\documentclass[12pt]{article}
\usepackage{amsmath,amsmath, amssymb}
\usepackage{graphicx}
\usepackage{natbib}
\usepackage{url} 

\newcommand{\blind}{0}

\usepackage{color,tikz,xcolor}
\usepackage{multirow}
\usepackage{pdflscape}
\newdimen\nodeDist
\usepackage[colorlinks,citecolor=blue,urlcolor=blue,filecolor=blue,backref=page]{hyperref}
\usepackage{caption, subcaption}
\usepackage{algorithm, algorithmicx,algpseudocode}
\usepackage{booktabs}

\usepackage{todonotes}
\newcounter{todocounter}

\newcommand{\res}{\mathrm{r}}
\newcommand{\X}{\mathbf{X}}
\newcommand{\x}{\mathrm{x}}
\newcommand{\y}{\mathrm{y}}
\newcommand{\yhat}{\widehat{y}}
\newcommand{\tr}{\text{tr}}
\newcommand{\te}{\text{te}}
\newcommand{\muvec}{\mathrm{\mu}}
\newcommand{\Tvec}{\mathrm{T}}
\newcommand{\indep}{\perp\!\!\!\perp}
\newcommand{\hypercube}{\mathcal{B}}
\newcommand{\Xtr}{\X^{\text{tr}}}
\newcommand{\Xte}{\X^{\text{te}}}
\newcommand{\restr}{\res^{\text{tr}}}
\newcommand{\reste}{\res^{\text{te}}}
\begin{document}

\def\spacingset#1{\renewcommand{\baselinestretch}%
{#1}\small\normalsize} \spacingset{1}


\if0\blind
{
  \title{\bf Local Gaussian process extrapolation for BART models with applications to causal inference}
  \author{Meijia Wang\hspace{.2cm}\\
    Arizona State University\hspace{.2cm}\\
    Jingyu He\hspace{.2cm}\\
    City University of Hong Kong\hspace{.2cm}\\
    P. Richard Hahn\hspace{.2cm}\\
    Arizona State University}
    \date{}
  \maketitle
} \fi

\if1\blind
{
  \bigskip
  \bigskip
  \bigskip
  \begin{center}
    {\LARGE\bf Local Gaussian process extrapolation for BART models with applications to causal inference}
\end{center}
  \medskip
} \fi

\bigskip
\begin{abstract}
Bayesian additive regression trees (BART) is a semi-parametric regression model offering state-of-the-art performance on out-of-sample prediction. Despite this success, standard implementations of BART typically provide inaccurate prediction and overly narrow prediction intervals at points outside the range of the training data. This paper proposes a novel extrapolation strategy that grafts Gaussian processes to the leaf nodes in BART for predicting points outside the range of the observed data. The new method is compared to standard BART implementations and recent frequentist resampling-based methods for predictive inference. We apply the new approach to a challenging problem from causal inference, wherein for some regions of predictor space, only treated or untreated units are observed (but not both). In simulation studies, the new approach boasts superior performance compared to popular alternatives, such as Jackknife+.
\end{abstract}

\noindent%
\vfill

\newpage
\spacingset{1.5} 

\section{Introduction}

Tree-based supervised learning algorithms, such as the Classification and Regression Tree (CART) \citep{breiman1984classification}, Random Forests \citep{breiman2001random}, and XGBoost \citep{chen2016xgboost} are popular in practice due to their ability to learn complex nonlinear functions efficiently. Bayesian Additive Regression Trees (BART, \cite{chipman2010bart}) is the most popular model-based regression tree method; it has been demonstrated empirically to provide accurate out-of-sample prediction (without covariate shift), and its Bayesian uncertainty intervals often out-perform alternatives in terms of frequentist coverage (see \cite{chipman2010bart, Kapelner2013PredictionWM}). XBART \citep{he2021stochastic} is a stochastic tree ensemble method that can be used to approximate BART models in a fraction of the run-time. Throughout the paper, we will refer to BART models but will use the XBART fitting algorithm.

\begin{figure}[h!]
    \centering
    \includegraphics[width = 0.5\textwidth]{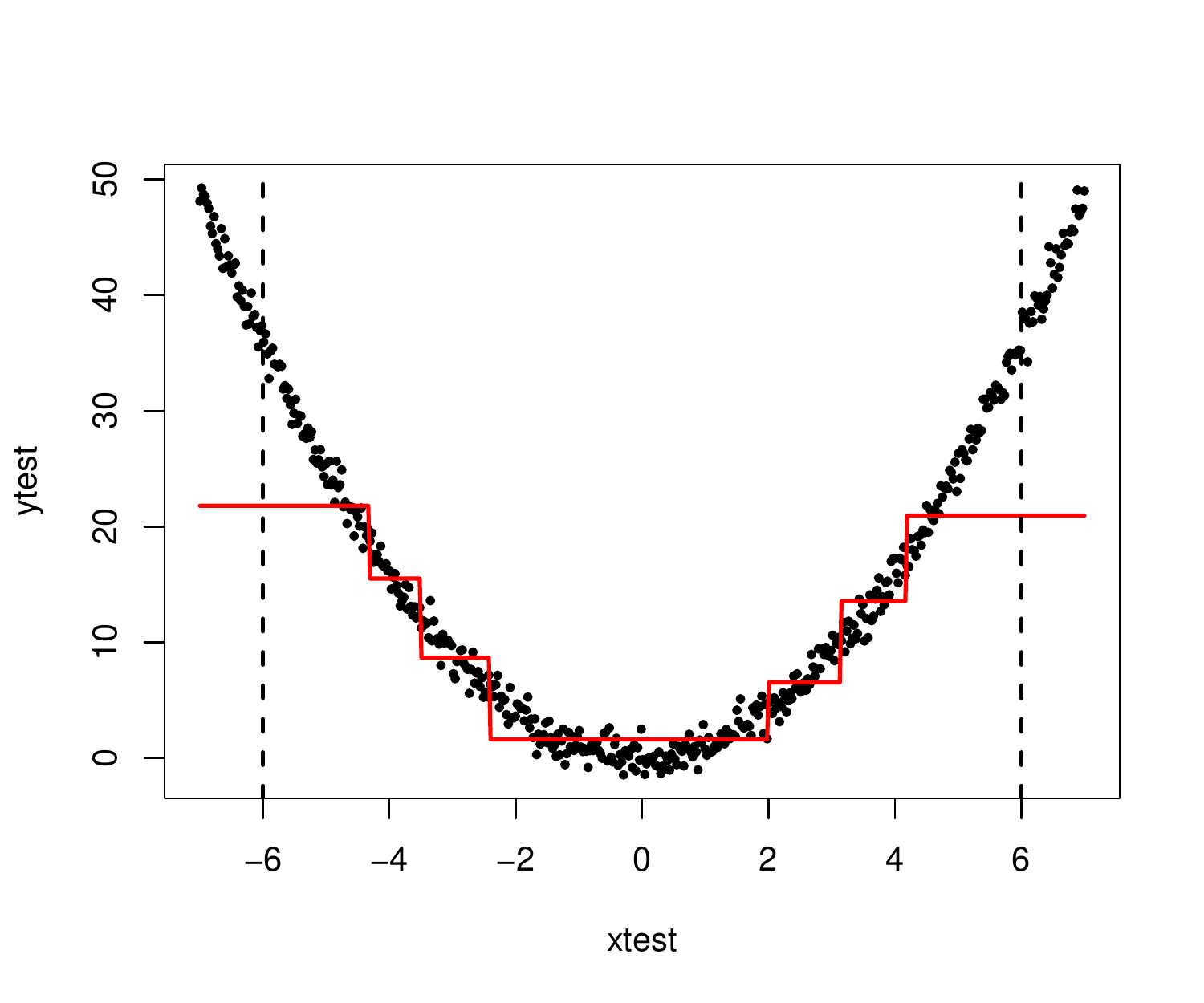}
    \caption{An example of the traditional tree extrapolation. The synthetic training data are black points. The fitted CART model is the step function in the solid line. The two vertical dashed lines are testing data to predict, which are outside the range of the training data. The traditional tree algorithms with constant leaf parameters extrapolate the prediction by a constant of the corresponding leaf node.}
    \label{fig:extrapolation}
\end{figure}

While tree-based methods frequently provide accurate out-of-sample predictions, their ability to extrapolate is fundamentally limited by their intrinsic, piecewise constant structure. Trees partition the covariate space into disjoint rectangular regions (leaf nodes) based on the observed covariates, then estimate a simple model (typically a constant) based on training data falling in the corresponding region. Consequently, to predict testing data outside the observed range of covariates in training, a naive tree regression model would extrapolate the constant in whatever leaf node that point falls in. Figure \ref{fig:extrapolation} illustrates the potential problem of extrapolation for the traditional tree models with constant leaf parameters in one-dimensional covariate space. If the testing data lie out of the range of the training, tree models have to extrapolate using the constant leaf parameter in the corresponding leaf node regardless of how far away the testing observations are from the nearest training point. 

In addition to poor point predictions, a flawed extrapolation model impairs posterior prediction intervals. Ideally, a BART model would predict essentially according to the prior at points far from the observed data; standard implementations instead never sample cutpoints beyond the training range and consequently dramatically understate the uncertainty in the unknown function in extrapolation regions. As a result, the coverage of predictive intervals from standard tree-based methods can be poor. While regression tree methods can be used in conjunction with resampling approaches to construct improved prediction intervals \citep{efron1981Jackknife, papadopoulos2008inductive, vovk2015cross,barber2021predictive}, such a strategy does not solve the intrinsically poor extrapolation of regression trees. 

This paper presents a novel approach to help tree methods extrapolate the exterior points and construct prediction intervals. The new method (XBART-GP) fits a Gaussian process (GP) at each leaf node of a BART tree and predicts exterior test points in the leaf by the associated GP model. Our approach differs from a treed Gaussian process \citep{gramacy2008bayesian} in that the local GP extrapolation does not impact the initial BART model fit. Intuitively,  model of \cite{gramacy2008bayesian} is a ``treed" GP, while our approach is a ``GPed" tree. The local GP extrapolation defines a novel posterior predictive distribution in terms of BART posterior samples and a GP model; it is not a new likelihood for inferring the trees themselves. Instead, the BART-inferred trees are used as inputs to the novel posterior predictive specification (cf. section \ref{sec:methodology}). Our proposed method shares a similar structure with the targeted smoothing BART (tsBART) model \citep{Starling2018tsBART, Starling2021heterogneous}, which fits a Gaussian process on a single targeted covariate in each tree leaf to introduce smoothness. In contrast, our approach fits the Gaussian process on a selection of covariates to extrapolate prediction.

In simulation studies, we compare our approach with recently proposed alternatives such as Jackknife+ \citep{barber2021predictive}. Our results illustrate that XBART-GP achieves nearly the desired coverage on out-of-range data while not substantially inflating the interval length. 

Additionally, our new extrapolation method is illustrated with an application to causal inference. When only treated (untreated) units are observable in a particular region of covariate space, inferring the untreated (treated) potential outcomes requires extrapolating the untreated (treated) response surface into that region. Such a scenario is said to violate the positivity condition necessary for identifying causal effects from observational data. Previous work on positivity violation includes \citep{DAMOUR2021644}, who focus on high-dimensional covariate spaces; \citep{nethery2019estimating}, who propose a spline model to extrapolate in non-overlap regions; and \citep{zhu2023addressing}, who construct a Gaussian process for the entire covariate space. In this paper, we apply the proposed local GP extrapolation approach to the Bayesian causal forest model \citep{hahn2020bayesian, nikolay2022}.


The rest of the paper is organized as follows. Section \ref{sec:background} reviews Gaussian processes, BART, Jackknife+, and Bayesian causal forests. Section \ref{sec:methodology} develops the new approach in detail and presents simulation studies demonstrating its advantages over the current state-of-the-art. Section \ref{sec:causal} adapts the method to the causal inference context and presents simulation results showing that local GP extrapolation of BART models can handle substantial violations of the overlap condition.

\section{Background}\label{sec:background}
\subsection{Notation}
We clarify notations throughout the paper. Let $\x = (x^{(1)}, \dots, x^{(p)})$ denote a $p$-dimensional covariate vector, or regressors. Capital letter $\X = (\x'_1, \dots, \x'_n)'$ denotes the $n\times p$ predictor matrix, and $\y = (y_1, \dots, y_n)$ is a vector of target values for supervised learning. First, we demonstrate our approach under the standard regression setting,
\begin{equation} \label{eq:regression}
    y_i = f(\x_i) + \epsilon_i, \quad \epsilon_i \sim N(0, \sigma^2),
\end{equation}
where we assume the residual term $\epsilon$ is a Gaussian noise term with mean zero and variance $\sigma^2$.

Let $C_{n, \alpha}(\x)$ denote the prediction interval for a test point $\x$ with confidence level $1-\alpha$ given $n$ predicted values. Notations $q^-_{n, \alpha / 2}\{y_i\}$ and $q^+_{n, \alpha / 2}\{y_i\}$ are the quantile functions for the $\lfloor \alpha / 2 (n+1) \rfloor$-th smallest value and the $\lceil(1-\alpha / 2)(n+1) \rceil$-th smallest value among a set of values $\{y_1, \dots, y_n\}$ respectively. 

\subsection{BART model and XBART algorithm}
The BART model \citep{chipman2010bart} assumes that the unknown function $f(\x)$ in the regression model \eqref{eq:regression} can be approximated by a sum of regression trees, i.e.
\begin{equation}
    f(\x) = \sum_{l=1}^L g(\x, \Tvec_l, \muvec_l),
\end{equation}
where $\Tvec_l$ represents a tree structure, which is a set of split rules partitioning the covariate space, and $\muvec_l$ is a vector of leaf parameters associated with the leaf nodes in tree $\Tvec_l$. 

\begin{figure}[!ht]
	\begin{subfigure}{171pt}
		\begin{center}
			\begin{tikzpicture}[
					scale=0.8,
					node/.style={%
							draw,
							rectangle,
						},
					node2/.style={%
							draw,
							circle,
						},
				]

				\node [node] (A) {$x_1\leq0.8$};
				\path (A) ++(-135:\nodeDist) node [node2] (B) {$\mu_{1}$};
				\path (A) ++(-45:\nodeDist) node [node] (C) {$x_2\leq0.4$};
				\path (C) ++(-135:\nodeDist) node [node2] (D) {$\mu_{2}$};
				\path (C) ++(-45:\nodeDist) node [node2] (E) {$\mu_{3}$};

				\draw (A) -- (B) node [left,pos=0.25] {no}(A);
				\draw (A) -- (C) node [right,pos=0.25] {yes}(A);
				\draw (C) -- (D) node [left,pos=0.25] {no}(A);
				\draw (C) -- (E) node [right,pos=0.25] {yes}(A);
			\end{tikzpicture}
		\end{center}

	\end{subfigure}
	\hfill
	\begin{subfigure}{171pt}

		\begin{center}
			\begin{tikzpicture}[scale=3]
				\draw [thick, -] (0,1) -- (0,0) -- (1,0) -- (1,1)--(0,1);
				\draw [thin, -] (0.8, 1) -- (0.8, 0);
				\draw [thin, -] (0.0, 0.4) -- (0.8, 0.4);
				\node at (-0.1,0.4) {0.4};
				\node at (0.8,-0.1) {0.8};
				\node at (-0.1, -0.1) {0};
				\node at (1, -0.1) {1};
				\node at (-0.1,1) {1};
				\node at (0.5,-0.2) {$x_1$};
				\node at (-0.3,0.5) {$x_2$};
				\node at (0.9,0.5) {$\mu_{1}$};
				\node at (0.4,0.7) {$\mu_{2}$};
				\node at (0.4,0.2) {$\mu_{3}$};
			\end{tikzpicture}
		\end{center}
	\end{subfigure}
	\caption{A regression tree partitions a two-dimensional covariate space (right) with two internal decision nodes (left). Each element of the partition corresponds to a terminal node in the tree with leaf parameter $\mu_l$.}\label{fig:tree_example}
\end{figure}
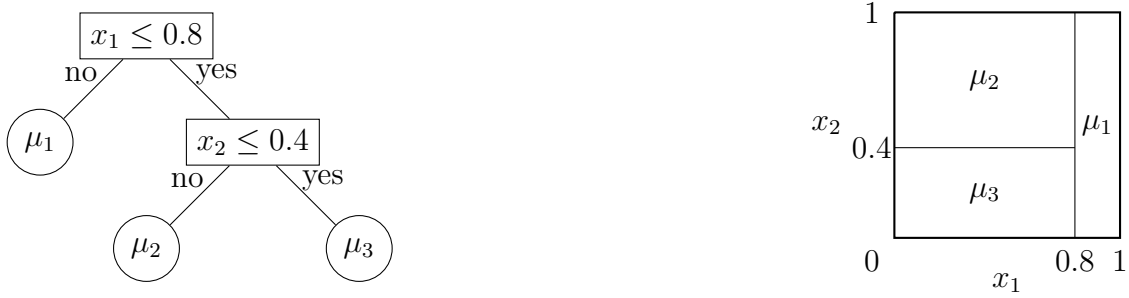

Figure \ref{fig:tree_example} illustrates a binary regression tree. The tree consists of a set of internal decision nodes that partition the covariate space to a set of terminal nodes (leaf nodes, say $\{\mathcal{A}_{1}, \cdots, \mathcal{A}_{B}\}$). The tree function $g(\x, \Tvec_l, \muvec_l)$ is essentially a step function due to the constant leaf parameter. It assigns point $\x$ with the leaf parameter of the leaf node that it falls into in tree $\Tvec_l$. 

Trees are known to be prone to overfitting due to their high flexibility. Thus, proper regularization is necessary to achieve good out-of-the-sample performance. BART assigns a regularization prior to the tree structure that strongly favors weak or small trees. The tree prior $p(\Tvec_l)$ specifies the probability for a node to split into two child nodes at depth $d$ to be
\begin{equation}
\alpha(1 + d)^{-\beta}, \quad \alpha \in (0, 1), \, \beta \in [0, \infty), \label{eq:tree_prior}
\end{equation}
which decreases exponentially as the tree grows deeper, implying strong regularization on the size of the tree. \cite{chipman2010bart} suggest choosing $\alpha = 0.95$ and $\beta = 2$. The prior of each leaf parameter $p(\mu_{lb})$ is assumed to be independent normal with variance $\tau$, i.e. $\mu_{lb} \sim N(0, \tau)$. The prior of the residual variance $\sigma^2$ is set to be $\text{inverse-Gamma}(a, b)$.

The ensemble of trees is fitted by Bayesian backfitting and MCMC sampling schemes. Let $\Tvec_{-l}$ denotes the set of all trees except $\Tvec_l$, and similary define $\muvec_{-l}$. Note that the conditional posterior $p(\Tvec_l, \muvec_l \mid \Tvec_{-l}, \muvec_{-l}, \sigma^2, y)$ depends on other trees and parameters only through the residuals
\begin{equation} \label{eq:resid}
    \res_l := y - \sum_{h\neq l}g(\x, \widehat{\Tvec}_h, \widehat{\muvec}_h) = g(\x, \Tvec_l, \muvec_l).
\end{equation}
Furthermore, one can integrate out the conjugate prior of $\muvec_l$ and derive the posterior of a tree $\Tvec_l$ in closed form, i.e.,
\begin{equation}
    p(\Tvec_l \mid \res_l, \sigma^2) \propto p(\Tvec_l) \int p(\res_l \mid \muvec_l, \Tvec_l, \sigma^2) p(\muvec_l \mid \Tvec_l, \sigma^2) d \muvec_l.
\end{equation}
This allows the conditional posterior $\Tvec_l \mid \res_l, \sigma^2$ and $\muvec_l \mid \Tvec_l, \res_l, \sigma^2$ to be drawn separately and sequentially for $l = 1, \dots, L$. After all trees are grown (we call it a sweep), the residual variance $\sigma^2$ is then sampled from the full conditional $\sigma^2 \mid \Tvec_1, \cdots, \Tvec_L, \muvec_1, \cdots, \muvec_L, y$. 

The original BART model \citep{chipman2010bart} draws trees from the posterior using a random walk Metropolis-Hastings (MH-MCMC) algorithm. Per iteration, the algorithm randomly proposes a single growing or pruning procedure to each tree and accepts or rejects it according to the MH ratios. 
Since the proposal in each iteration only makes a small modification, one can expect it is not efficient for the MH-MCMC algorithm to explore the posterior model space. The experiments in \citep{chipman2010bart} typically use $200$ burn-in steps and $1000$ iterations with $200$ trees. Furthermore, the algorithm does not scale well with large sizes of data observations or covariates. 

XBART \citep{he2019xbart, he2021stochastic}, shortened for accelerated Bayesian additive regression trees, is a stochastic tree ensemble method inspired by BART. The essential idea of XBART is to avoid the expensive MH-MCMC computation but grow completely new trees recursively (like CART). At each node, the model considers a similar set of cutpoints as BART and uses the marginal likelihood to sample split criteria proportionally. Therefore it is much more efficient to explore the posterior of BART and scales well. Our approach is designed for both the framework of XBART and BART, while we will demonstrate the performance using XBART for computational efficiency.

To obtain the posterior prediction interval from XBART, we take the sampled trees as draws from a standard Bayesian Monte Carlo algorithm. Suppose the algorithm draws $S$ sweeps with $S_0$ burn-out iterations and $L$ trees per forest. The predicted posterior mean and variance at iteration $s$ are defined as $\widehat{f}_s(\x) = \sum_{l=1}^L g_l(\x, \widehat{\Tvec}_l^{(s)}, \widehat{\mu}_l^{(s)})$ and $\widehat{\sigma}_s^2$ respectively. The final prediction of target value $\widehat{y}_s$ is sampled from $N(\widehat{f}_s(\x), \widehat{\sigma}_s^2)$. Furthermore, the prediction interval of target level $1-\alpha$ is given by quantiles of the predictions as follows
\begin{equation}
\widehat{C}^{\text{XBART}}_{S,\alpha}(\x) = \left[\widehat{q}^-_{S,\alpha / 2}\{\yhat_s\}, \widehat{q}^+_{S,\alpha / 2}\{\yhat_s\}\right] \label{eq:xbart_pi}.
\end{equation}
Note that this definition of posterior prediction interval is based on the prediction of every single tree with a constant leaf parameter, i.e., it still suffers from the extrapolation problem when the testing data is out of the range of the training. 

\subsection{Gaussian process regression}
This section reviews Gaussian process regression, a critical ingredient of the proposed XBART-GP extrapolation. Gaussian process regression is a non-parametric kernel-based Bayesian probabilistic model. It can interpolate and extrapolate as the covariance defines the model's behavior over its full domain. Essentially, the specified covariance function characterizes the linear dependence of the response at pairs of points as a function of some measure of distance between those points in covariate space. For a textbook treatment of the Gaussian process, see \cite{rasmussen2003gaussian}. 

More specifically, Gaussian process regression assumes $\{ f(\x), \x \in \mathbb{R}^p \}$ is a set of random variables that follows the Gaussian process with some mean function $\mu(\x)$ and covariance function $k(\x, \x^*)$. Let $(\X, \y)$ denote a set of training data, and $\X^*$ denote the input of testing data. Following the standard regression setting in Equation \ref{eq:regression}, the joint distribution of function values $f(\X^*)$ and $\y$ is given by 
\begin{equation}
    \begin{pmatrix}
    f(\X^*) \\ \y
    \end{pmatrix}
    \sim
    N\left( 
    \mu \begin{pmatrix}
    \X^* \\ \X
    \end{pmatrix}, 
    \begin{pmatrix}
    \Sigma_{\X^* \X^*}, &\Sigma_{\X^* \X}\\
    \Sigma_{\X \X^*}, &\Sigma_{\X\X} + \sigma^2\mathbf{I}
    \end{pmatrix}
    \right)
\end{equation}
where $\mu(\cdot)$ is a regression function, and the covariance matrices are calculated by pre-specified kernel function $k(\x, \x^*)$ 

The prediction of the outcome $f(\X^*)$ can be drawn from the conditional distribution $f(\X^*) \sim N(\mu_{\X^*\mid\X}, \Sigma_{\X^*\mid\X})$ with conditional mean and covariance:
\begin{equation}
    \mu_{\X^* \mid \X} = \mu(\X^*) + \Sigma_{\X^*\X} \left[\Sigma_{\X\X} + \sigma^2\mathbf{I}\right]^{-1} (\y - \mu(\X) )
\end{equation}
\begin{equation}
    \Sigma_{\X^*\mid\X} = \Sigma^2_{\X^*\X^*} - \Sigma_{\X^*\X}\left[\Sigma_{\X\X} + \sigma^2\mathbf{I}\right]^{-1} \Sigma_{\X\X^*}.
\end{equation}

\section{Local GP extrapolation for BART}\label{sec:methodology}

Traditional BART prediction intervals in Equation (\ref{eq:xbart_pi}) provide good coverage with tight intervals on many data generating processes \citep{he2021stochastic} with exchangeable train and test data. However, predictions are less reliable when the train and test sets differ substantially in terms of their support. The intuition behind our method is to use the tried-and-true BART predictions (and intervals) for prediction points within the range of the training data but to incorporate Gaussian process extrapolation for points well outside this support. Therefore, the first step in describing the new method is to formally define the concepts of exterior points in the context of regression trees. Specifically, we will define extrapolation points \textit{locally} in terms of the regression trees: a test point $\x'$ is an \textit{exterior} point if and only if it is outside the range of the training data in the leaf node it falls in. The basic strategy of the new method is to use the leaf-specific training data to extrapolate such exterior points using a local Gaussian process. Combining these local GP predictions across trees (and across posterior samples) constitutes the main technical content of the new method, details of which are below.

Before getting into those details, a big-picture explanation may be helpful. Typically, a Bayesian posterior predictive distribution is
$$f(\tilde{y} \mid y_{1:n}, \X_{1:n}) = \int f(\tilde{y} \mid \theta) \pi(\theta \mid y_{1:n}, \X_{1:n}) d\theta ,$$
which conveys the idea that future and past data are conditionally independent given the model parameters. Here, we explicitly deviate from this formulation, but only for those future points that are exterior (defined as a function of the model parameters). For such a point, our posterior predictive is defined as 

$$f(\tilde{y} \mid y_{1:n}, \X_{1:n}) = \int f(\tilde{y} \mid \theta, y_{1:n}, \X_{1:n}) \pi(\theta \mid y_{1:n}, \X_{1:n}) d\theta
$$
where the predictive distribution explicitly involves {\em both} the training data and model parameters (trees and leaf means in the case of BART). In this sense, the proposed approach does ``use the data twice'' but not in the construction of the posterior. Rather, our procedure is a way of combining the orthodox BART posterior with an intentionally and explicitly distinct posterior predictive model for extrapolation points. Although this is an uncommon approach that violates so-called ``diachronic coherence'' \citep{skyrms2006diachronic}, it is in no sense illicit in that it merely amounts to specifying a user-defined predictive distribution that takes BART posterior samples as inputs. From this perspective, describing the method amounts to providing a definition of the predictive kernel, $f(\tilde{y} \mid \theta, y_{1:n}, \X_{1:n})$. Specifically, this will be a Gaussian process with a covariance kernel defined in terms of the trees of the BART model.


\subsection{Notation}
Consider a fitted BART forest $\{\widehat{\Tvec}_{l}\mid 1 \leq l \leq L\}$ with $L$ trees. Let $\{\mathcal{A}_{l1}, \cdots, \mathcal{A}_{lB_l}\}$ denote the covariate space partitioned by the $l$-th tree $\widehat{\Tvec}_l$ with $B_l$ leaf nodes, and $\widehat{\sigma}^2$ for the estimated residual variance. For the $l$-th tree, suppose the test point $\x$ falls in a leaf node $b$ with covariate space $\mathcal{A}_{lb}$. Let the training data that falls in this leaf node and its residuals at the current tree (defined by Equation \eqref{eq:resid}) be noted as $\Xtr$ and $\restr$. Similarly, $\Xte$ and $\reste$ denote the exterior testing data in the leaf node $b$ and its to-be-predicted residuals.

Note that the tree partitioned leaf node $\mathcal{A}_{lb}$ can extend to infinity if they are at the boundary. Specifically, the range of subset of training data $\Xtr$ falling in the $b$-th leaf node of the $l$-tree forms a $p$-dimensional hypercube $\hypercube_{lb}$, which is a subset of $\mathcal{A}_{lb}$. If the testing point $\x$ falls within the range of training data, i.e., $\x \in \hypercube_{lb}$, we consider it as an interior point, and its prediction follows the standard XBART model. Otherwise, if $\x \in \{\mathcal{A}_{lb} \setminus \hypercube_{lb}\}$, it is an exterior point, and their predictions require extrapolation. Notably, this definition is local. Testing data can be exterior for some of the trees and interior for others in the XBART forest.

\subsection{Defining the GP model}
To extrapolate the prediction of XBART, we assume that the exterior testing data $\Xte$ and training data $\Xtr$ in the same leaf node form a Gaussian process. The joint distribution of the training and testing data in a leaf node is given by
\begin{equation}
    \begin{pmatrix}
    \reste\\ \restr
    \end{pmatrix}
    \sim N\left( 
    \mu \mathrm{J},
    \begin{pmatrix}
    \mathbf{\Sigma}_{\Xte \Xte}, &\mathbf{\Sigma}_{\Xte \Xtr}\\
    \mathbf{\Sigma}_{\Xtr \Xte}, & \mathbf{\Sigma}_{\Xtr \Xtr} + \frac{1}{L}\widehat{\sigma}^2 \mathbf{I}
    \end{pmatrix}
    \right)
\end{equation}
where $\mu$ is the leaf parameter, $\mathrm{J}$ is a column vector of ones, and $\mathbf{I}$ is an identity matrix. By definition, the variance of the response $y$ is assumed to be $\sigma^2$. Here we assume that the residuals in each tree contribute equally to the total variance. 

To reflect the relative distance among covariates, we define the covariance function of the Gaussian process as the squared exponential kernel 
\begin{equation} \label{eq:cov_func}
k(\x, \x') = \tau \exp \left(- \theta \sum_{i=1}^p\frac{(x_i - x'_i)^2}{2\delta_i^2}\right)
\end{equation}
where $\delta_i$ is the range of the $i$-th variable $x_i$ in a leaf node, $\theta$ controls the smoothness and $\tau$ determines the scale of the function.
$\x = (x_1,\cdots, x_p)$ and $\x' = (x_1',\cdots,x_p')$ are the vector of two data points. In our experiments, we use $\theta = 0.1$ and set $\tau$ to be the variance of observed responses divided by the number of trees.

The conditional distribution of the residuals $\reste$ is given by
\begin{equation} \label{eq:conditional}
\begin{aligned}
\mathrm{\mu}_{\Xte \mid \Xtr} &= \mu \mathrm{J} + \mathbf{\Sigma}_{\Xte \Xtr} \left[\mathbf{\Sigma}_{\Xtr \Xtr} + \frac{1}{L}\widehat{\sigma}^2 \mathbf{I} \right]^{-1}(\restr - \mu  \mathrm{J})\\
\mathbf{\Sigma}_{\Xte \mid \Xtr} &= \mathbf{\Sigma}_{\Xte} - \mathbf{\Sigma}_{\Xte \Xtr} \left[\mathbf{\Sigma}_{\Xtr \Xtr} + \frac{1}{L}\widehat{\sigma}^2 \mathbf{I} \right]^{-1} \mathbf{\Sigma}_{\Xtr \Xte}.
\end{aligned}
\end{equation}
The prediction for the residuals of the exterior testing data associated with this leaf node is drawn from the conditional normal distribution $\reste \sim N(\mathrm{\mu}_{\Xte\mid \Xtr}, \mathbf{\Sigma}_{\Xte \mid \Xtr})$ rather than the fixed leaf parameter of the tree, which is used for predicting interior points. 


\subsection{Algorithms}
It can be computationally intensive to find the corresponding training data and calculate the covariance matrix for each test data observation. Following the fast algorithm GrowFromRoot introduced in \cite{he2019xbart}, we propose an efficient method that pinpoints the exterior testing data and its neighborhood training data in the same leaf node by recursively partitioning the training and testing data at the same time. The procedures are summarized in Algorithm \ref{alg:predict} and \ref{alg:PFR}. 

\begin{algorithm*}[!ht]
	\small
	\caption{Prediction by GP extrapolation}\label{alg:predict}
	\begin{algorithmic}[1]
	\Procedure{predict}{$\Xte, \Xtr, \y_{\tr}, \{\Tvec_{sl} \}, \{\mathrm{\mu}_{sl}\}, L, S$} \\
		\textbf{output} $S$ Monte Carlo draws of predicted mean $\{\widehat{\y}^{\te}_s\}$ for input data $\Xte$
		\State Create empty vectors $\{\restr_{sl}\}$ and $\{\reste_{sl}\}$ to store the predicted residuals.
		\State $\res \leftarrow \y - \bar{\y}$ \Comment{Initialize full residuals for training set $\X$.}
		\State $\restr_{0l} \leftarrow \bar{\y} / L $ \Comment{Initialize predicted residuals for the first iteration.}
		\For{$s$ in 1 to $S$}
		\For{$l$ in 1 to $L$}
		
		\State Create an empty vector $\mathrm{v}$ to store split variables.
		\State $\res \leftarrow \res - \restr_{s-1,l}$ \Comment{Update partial residuals.}
		\State PFR$\left(\res, \Xtr, \restr_{sl}, \Xte,  \reste_{sl}, \Tvec_{sl}, \mathrm{\mu}_{sl}, \mathrm{v}, {\tt root\_node}\right)$  \Comment{Call Algorithm 2.}
		\State $\res \leftarrow \res + \restr_{sl}$ \Comment{Update full residuals.}
		\EndFor
		
		\State $\widehat{\y}_{s}^{\text{te}} \leftarrow \sum_{l=1}^L \res_{sl}^{\text{te}}$ \Comment{Calculate predicted mean from the $s$-th iteration.}
		
		\EndFor
		\State Return predicted means $\{ \widehat{\y}_s^{\text{te}} \}$.
		\EndProcedure
	\end{algorithmic}
\end{algorithm*}

\begin{algorithm*}[h!]
	\small
	\caption{PredictFromRoot}\label{alg:PFR}
	\begin{algorithmic}[1]
		\Procedure{PFR}{$\res,  \Xtr, \restr,  \Xte, \reste$, $T_l$, $\mathrm{\mu}_l$, $\mathrm{v}$, {\tt node}} \\
		\textbf{outcome} Predict residuals $\reste$ for testing covariates $\Xte$ using Gaussian process. 
		\If{{\tt node} is leaf node}
        \State Define hypercube $\hypercube_{lb}$ based on the range of $\Xtr$.
		\State Define active variables $\mathrm{v}_b$ based on $\Xte$ and $\hypercube_{lb}$. \Comment{If there exists a testing point $\x \in \Xte$ such that $\x$ is out of the range of $\hypercube_{lb}$ on that variable, then it is considered as an active variable.}
		\State Sample a subset of training data $\Xtr_b$ and $\restr_b$ of size $100$ with only variables in $\mathrm{v}_b$.
		\State Get exterior testing data $\Xte_b$ with only variables in $\mathrm{v}_b.$
		\State  Get covariance matrix $\Sigma_{\Xtr_b \Xtr_b}$ and its inverse $\Sigma^{-1}_{\Xtr_b \Xtr_b}$ for the Gaussian process.
		\State Get conditional mean and variance $\mu_{\Xte_b\mid \Xtr_b}$, $\sigma_{\Xte_b \mid \Xtr_b}$ from Equation \eqref{eq:conditional}.
		\State $\restr \leftarrow \mu_{lb}$ \Comment{Assign leaf parameter $\mu_{lb}$ as predicted residuals.} 
		\State $\reste_b \leftarrow \mu_{\Xte_b\mid \Xtr_b} + \Sigma_{\Xte_b\mid \Xtr_b} \mathrm{z} $
		\Comment{Sample predictions from Gaussian process; $\mathrm{z}$ is a vector of standard normal variables.}
		\For{$i$ in 1 to $N_t$} \Comment{$N_t$ is the number of the testing data.}
		\If{$\Xte[i] \in \hypercube_{lb}$}
		\State $\reste[i] \gets \mu_{lb}$ \Comment{Assign leaf parameter as predicted residual.}
		\Else
        \State $\reste[i] \leftarrow \reste_b[i]$ \Comment{Assign prediction from the Gaussian process.}
		\EndIf
		\EndFor
		\Else \Comment{Split until reaching leaf nodes.}
		\State Get split variable $v$ and cutpoint $c$. Add $v$ to splitting variables $\mathrm{v}$.
		\State Shift $\res,  \Xtr, \restr,  \Xte, \reste$ into left and right parts according to $v$ and $c$.
		\State PFR$\left(\res_{\mbox{left}},  \Xtr_{\mbox{left}}, \restr_{\mbox{left}},  \Xte_{\mbox{left}}, \reste_{\mbox{left}}, T_l, \mu_l, \mathrm{v}, {\tt left\_node}\right)$
		\State PFR$\left(\res_{\mbox{right}},  \Xtr_{\mbox{right}}, \restr_{\mbox{right}},  \Xte_{\mbox{right}}, \reste_{\mbox{right}}, T_l, \mu_l, \mathrm{v}, {\tt right\_node}\right)$
		\EndIf
		\EndProcedure
	\end{algorithmic}
\end{algorithm*}

Certain algorithm design choices to boost performance and efficiency are worth emphasizing. First, we partition the covariates into \textit{split variables} $\mathrm{v}$ and \textit{active variables} $\mathrm{v}_b$. Split variables appear along the decision path from the root to a leaf node. Active variables $\mathrm{v}_b$ are a subset of split variables that satisfies the following condition: If there exists a testing point $\x \in \Xte$ such that $\x$ is outside the range of $\hypercube_{lb}$ on that variable in leaf node $b$, then it is considered an active variable. Using only active variables in defining the GP covariance matrix ensures that the extrapolation depends only on those variables, which is intuitive and improves time efficiency. Second, when the training data sample size is large, we subsample training data when performing the GP extrapolation. Third, to avoid the identification of exterior points being misled by any outliers, we define the hypercube $\mathcal{B}_{lb}$ in each leaf node by the $95\%$ quantile of the training data.

\subsection{Simulation studies} \label{sec:sim}
This section illustrates the performance of XBART-GP and compares it to various prediction inference methods in regression modeling. We carefully construct data generating processes with covariate shifts and test the model's predictive ability on interior and exterior data points separately. 

Specifically, we compare our proposed method with the latest frequentist approach Jackknife+ \citep{barber2021predictive} and its cross-validation version (CV)+. For any regression model, Jackknife+ uses the leave-one-out procedure to train the model repeatedly and obtain out-of-sample residuals to construct the predictive interval. Similarly, CV+ is a computationally efficient version of Jackknife+ that uses k-fold cross-validation to obtain the model's out-of-sample uncertainty. We apply the two methods on XBART and Random Forest respectively, as the baseline methods.

For both XBART and XBART-GP, we use the following hyperparameters {\tt num\_trees} $= 20$,  {\tt num\_iterations} $=100$, ${\tt Nmin = 20}$, $\tau = \text{Var}(\y)/${\tt num\_trees} (where $\text{Var}(\y)$ is the variance of the response variable in the training set). We choose $\theta_{gp} = 0.1$ and $\tau_{gp} = \tau$ to be the smoothness and scale parameters of the kernel function for XBART-GP.

When applying Jackknife+ and CV+ to XBART, we pick the same number of trees and $\tau$ value, except that the number of iterations is reduced to $100$ to reduce the running time of the two approaches. For simulations of Jackknife+ and CV+ on Random Forest, called Jackknife+ RF and CV+ RF in the rest of the paper, we use the default settings provided in the {\tt scikit-learn} \citep{scikit-learn} package in \texttt{Python}.

The real signal is drawn from four challenging functions, $f$, listed in Table \ref{tab:truef}. The response variable $y$ is generated independently from $y = f(\x) + \epsilon$ with Gaussian noise $\epsilon$. In all cases, we generate $200$ data points with covariates $x_j \overset{\text{iid}}{\sim} N(0, 1)$ for $j  = 1, \dots, d=10$ as the training set and another $200$ data points with $x_j \overset{\text{iid}}{\sim} N(0, 1.5^2)$ as the testing set. 

\begin{table}[!ht]
\small
\def\arraystretch{1}
\centering 
\begin{tabular}{ll} 
\toprule
Name &Function \\
\hline
Linear &     $ \x^t \mathrm{\gamma} $;\; $ \gamma_j = -2+ \frac{4(j-1)}{d-1} $\\ 

Single index & $10\sqrt{a} + \sin{(5a)}$;\;$a=\sum_{j=1}^{10} (x_j - \gamma_j)^2$;\; $\gamma_j = -1.5+ \frac{j-1}{3}$.\\


Trig + poly & $5\sin(3x_1)+2x_2^2 + 3x_3x_4 $\\

Max & $\max(x_1,x_2,x_3) $\\ 
\bottomrule
\end{tabular}
\caption{Four true $f$ functions} 
\label{tab:truef}
\end{table}

Since the definition of exterior point for XBART-GP depends on the tree's structure and varies by leaf nodes, we simplify the concept in this section to evaluate performance on in- and out-of-range data. In our simulation studies, we consider any testing data outside the training data range as exterior points. Under the above settings, approximately half of the testing data is interior, and half is exterior. The performance is then reported on interior and exterior points separately. We repeat the experiments on each data generating process $10$ times and calculate the average empirical probability of coverage and the average interval length with coverage level $1 - \alpha = 0.9$. 

Simulation studies and time efficiency comparison on XBART-GP and its baselines were conducted in Python $3.8.10$ on a Linux machine with Intel(R) Core(TM) i7-8700K CPU @ 3.70GHz processor and 64GB RAM; eight cores were used for parallelization whenever it was applicable. 

\begin{figure}[!ht]
\centering
\begin{subfigure}{379pt}
 \subcaption[c]{Linear}
 \includegraphics[width=379pt]{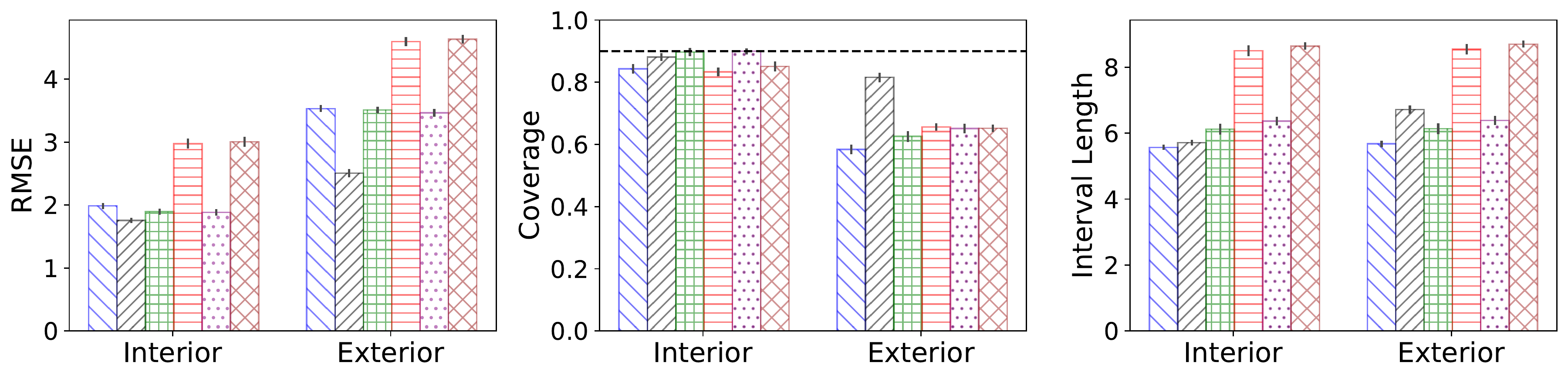}
\end{subfigure}
\centering
\begin{subfigure}{379pt}
 \subcaption[c]{Single Index}
 \includegraphics[width=379pt]{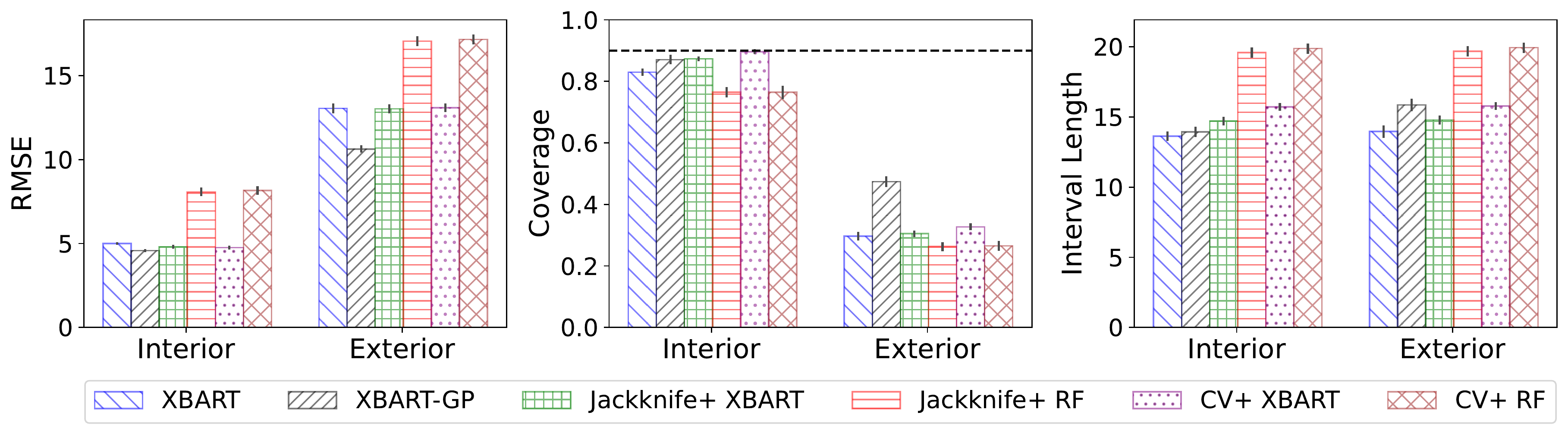}
\end{subfigure}
\caption{Visualization of simulation results in root mean square error (RMSE), interval coverage (Coverage), and interval length for interior and exterior points for linear and single index functions. The horizontal dash line indicates the $90\%$ target coverage level. The vertical lines on top of each bar shows the deviation from repeated experiments.}
\label{fig:simulation}
\end{figure}

Figure \ref{fig:simulation} visualizes the simulation results on linear and single index functions. The full results of four functions are summarized in the Appendix Figure \ref{fig:simulation2} and Table \ref{tab:simulation}. This figure compares the root mean square error (RMSE) of point prediction, coverage rate, and interval length of various approaches. Note that the RMSE of Jackknife+ and CV+ are evaluated based on the average leave-one-out prediction and cross-validated prediction, respectively.

We notice that XBART-GP has the smallest RMSE of point prediction across all functions on both interior and exterior points. For interior data, our approach is the only one that delivers nominal coverage while the interval length is not too large. Furthermore, it also has substantially greater prediction coverage than the baseline methods on exterior points. It is worth noting that XBART-GP achieves better coverage without inflating the prediction interval too much. Indeed, the interval length of XBART-GP is just slightly larger than that of Jackknife+XBART or CV+XBART but is still smaller than that of Jackknife+RF or CV+RF. 

Combining Jackknife+ and CV+ with XBART, the two methods produce competitive RMSE, coverage, and interval length compared to XBART itself on interior points. However, the prediction inference methods also fail to extrapolate as the in-sample residuals can not inform uncertainty on out-of-range data.

\subsubsection{Time efficiency comparison}
We then evaluate the performance and time efficiency of XBART-GP and other baselines for increasing sample size.

The training and testing covariates are generated from the same distributions as in the previous section, with testing covariates having a slightly larger variance. We generate the response variable $Y$ with the linear function in Table \ref{tab:truef} and the standard normal error term. We estimate the average coverage, interval width, and computational time of all $6$ methods on with sample size $n = n_t = 50, 100, 150, 200, 300, 500$ over $10$ independent trials for each sample size. The target coverage level is $1 - \alpha = 0.9$.
 
\begin{figure}[!ht]
 \centering
 \includegraphics[width=379pt]{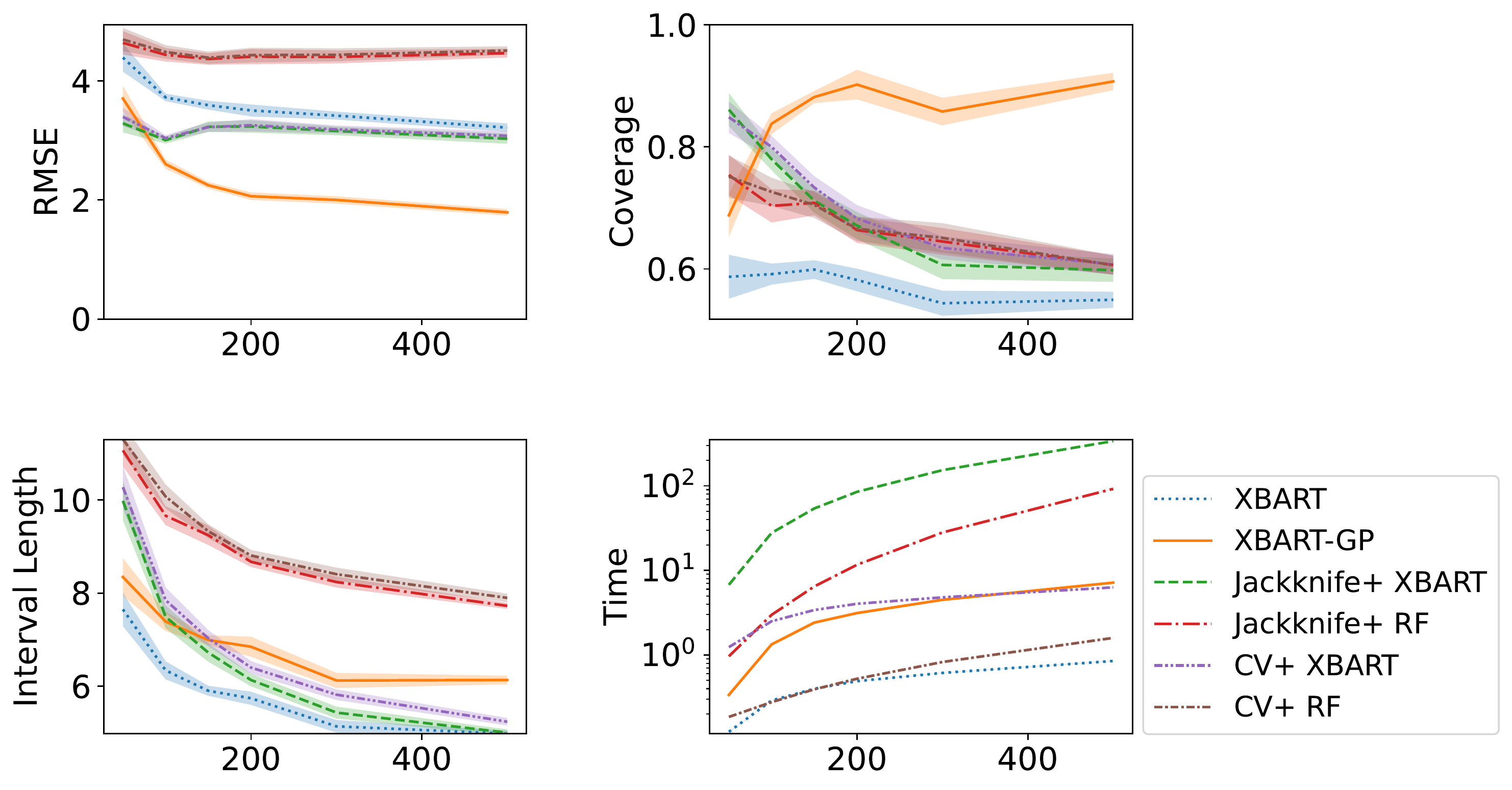}
\caption{Average RMSE, coverage, interval length on exterior points, and running time (in seconds) at $90\%$ level for all methods over $10$ independent trials with increasing sample size from $50$ to $500$. Time in seconds is presented in logarithm scale.}
\label{fig:time_exterior}
\end{figure}

Figure \ref{fig:time_exterior} illustrates the performance comparison of XBART-GP and other baseline methods on exterior points with increasing sample size. The performance on interior points is similar and is presented in Appendix Figure \ref{fig:time_interior}. Note that the proportion of exterior points generated varies according to the training size. Empirically, the percentage of exterior points decreases from $80\%$ to $40\%$ when the sample size increases from $50$ to $500$ on $10$-dimensional covariates. The x-axises in Figure \ref{fig:time_exterior} only reflect the sizes of the training sets. The running time is reported as evaluating all testing data, including interior and exterior points.

On exterior points, XBART-GP has the smallest RMSE of point prediction and a stable coverage close to $90\%$ as the sample size grows. The coverage of other baseline methods drops significantly on a larger sample size. The interval of XBART-GP is slightly wider than XBART but still outperforms random forest approaches.  

XBART and CV+ approaches are the most efficient in terms of running time. The run-time of XBART-GP increases almost linearly with the amount of training and testing sample size, but it is worth the time when the distribution of the test set shifts from the training set.



\section{Application of local GP extrapolation on causal inference}\label{sec:causal}
\subsection{Background}\label{sec:XBCF}
A motivating example of an applied extrapolation problem occurs in treatment effect estimation, in particular, when one of the necessary assumptions is violated under the potential outcome framework \citep{imbens2015causal} for estimating the treatment effect. 

Let $Y$ denote the response variable and $Z$ denote the binary treatment variable with $Z_i = 1$ representing the $i$-th observation being treated or $Z_i = 0$ as being in the control group. The potential outcomes under treatment and control are denoted as $Y_i(1)$ and $Y_i(0)$ for the $i$-th unit, respectively. The observable outcome can be expressed as $Y_i = Z_i Y_i(1) + (1-Z_i) Y_i(0)$. The treatment effect on an individual $\x_i$ is defined as $\tau(\x_i) = Y_i(1) - Y_i(0)$.

Most causal inference methods using the potential outcome framework also rely on the following assumptions: 
\begin{enumerate}
    \item Stable Unit Treatment Value Assumption (SUTVA); 
    \item Ignorability assumption:  $Y_i(1), Y_i(0) \indep Z_i \mid \X_i$; 
    \item Positivity assumption: $0 < P(Z_i = 1\mid \x_i) < 1$. 
\end{enumerate}

The positivity assumption is also known as the overlap assumption. In this study, we differentiate the covariate space into two parts, the overlap region where the positivity assumption is satisfied, and the non-overlap region where it is not. As per \cite{Zhu2021PositivityViolation}, the violation of positivity assumptions can be categorized into two categories, structural and practical violation.

Structural violations occur when it is impossible for individuals with specific covariate values to receive treatment, leading to the treatment effect for these individuals or covariate values being irrelevant. One solution is to "trim" the population, as described in \citep{ho_imai_king_stuart_2007, petersen2012, crump2009}, which allows for the estimation of the treatment effect only in the overlap region (where the positivity assumption holds).

Practical violation refers to situations where a subset of covariate values is only observed in the treatment or control group due to selection bias or a limited sample size. Nevertheless, the average treatment effect (ATE) over the entire population and the conditional average treatment effect (CATE) for specific covariate values are still of interest \citep{yang_ding_2018}. In these cases, the strategy is to perform model-based extrapolation, extending an estimated response surface into regions where no data was observed. 

Recently, a few extrapolation methods have been developed to estimate the ATE of the entire population with poor overlap. \cite{fanli2018a} and \cite{fanli2018b} proposed using overlap weights to balance the covariates for estimating the ATE. \citep{nethery2019estimating} proposed a way to define overlap and non-overlap regions based on the propensity score. They then use BART to estimate the treatment effect in the overlap region and use a spline model to extrapolate the treatment effect in the non-overlap region. \cite{zhu2023addressing} addressed this problem by modeling the treatment effect with the Gaussian process model, which by its nature can extrapolate based on the covariate kernel. 

\subsection{Local GP extrapolation with Bayesian causal forest}
In this section, we explore extrapolating CATE in the region of non-overlap by applying the local GP extrapolation method on the Accelerated Bayesian Causal Forest (XBCF) model \citep{nikolay2022}. First, we briefly review the structure of the XBCF model. Next, we formally define the overlap and non-overlap area in the ensemble trees framework. Then we describe the modifications we have made to integrate the local GP algorithm for extrapolating the treatment effect estimated by XBCF.

XBCF is a recent approach for estimating heterogeneous treatment effect based on Bayesian Causal Forest \citep{hahn2020bayesian}. The model expresses the observable outcomes as 
\begin{equation} \label{eq:xbcf}
\begin{aligned}
   y_i &= a \mu(\x_i, \widehat{\pi}_i)) + b_{z_i} \tau(\x_i) + \epsilon_i, \quad  \epsilon_i \sim N(0, \sigma^2_{z_i}),\\ 
   &a\sim N(0,1), \quad b_0, b_1\sim N(0, 1/2)
\end{aligned}
\end{equation}
where $\widehat{\pi}_i$ is the estimated propensity score. Function $\mu(\x_i, \widehat{\pi}_i)$ aims to capture the prognostic effect and function $\tau(\x_i)$ captures the treatment effect. Both functions are modeled by a sum of XBART trees, respectively. 

Under the derived split criterion of XBCF, the treatment trees in $\tau(\x_i)$ stop splitting when the leaf nodes only consist of either the treatment group or the control group. However, the prognostic forest $\mu(\x_i, \widehat{\pi}_i)$ is not affected by the positivity violation as it does not depend on the treatment variable $z$. In this case, both estimators will be biased when the positivity assumption is violated. However, the tree structure effectively captures the heterogeneity in the overlap region, making it ideal for extrapolating causal effects in the non-overlap region using the Gaussian process model in the leaf nodes.

The model relies on the three basic assumption in the potential outcome framework describe in the previous section. However, when the positivity assumption is violated in practice, we can not make firm conclusions about ATE or CATE in the non-overlap region. To mitigate this limitation, we make the assumption that there is no abrupt change in the treatment effect function in both overlap and non-overlap regions, which allows us to make inferences about CATE using the extrapolated predictive interval obtained from local Gaussian processes.

We apply the local GP technique on the XBCF model to extrapolate the treatment effect in non-overlap regions as follows. Given a regression tree, consider a leaf node $b$ with local covariate subspace $\mathcal{A}_{\X}$, let matrix $\Xtr$ denote the training data that falls into the leaf node, which can be separated into a treatment group $\Xtr_T$ and a control group $\Xtr_C$ based on the corresponding treatment variable $z$. The range of covariates in the treated and the control form two hypercubes, here we denote as $\mathcal{B}_T$ and $\mathcal{B}_C$ respectively. The overlap region in the leaf node is the intersection of $\mathcal{B}_T$ and $\mathcal{B}_C$, or $\mathcal{B}_O = \mathcal{B}_T \cap \mathcal{B}_C$. Consequently, the non-overlap area is $\mathcal{B}_{NO} = \mathcal{A}_{\X} \setminus \mathcal{B}_O$.

It's important to note that in each treatment tree, the residuals in the non-overlap region are biased in direction relative to the true treatment effect. This is because of the treatment trees' no-splitting behavior. As a result, if testing data is included in the Gaussian process, it will be extrapolated in the opposite direction. Therefore, different from XBART-GP, we assume that the testing data in the non-overlap area $\Xte_{\text{NO}}$ share a joint distribution with only the overlap training data $\Xtr_{\text{O}}$, namely
\begin{equation}
    \begin{pmatrix}
    \reste_{\text{NO}} \\ \restr_{\text{O}}
    \end{pmatrix}
    \sim N\left( 
    \mu \mathrm{J},
    \begin{pmatrix}
    \mathbf{\Sigma}_{\Xte_{\text{NO}} \Xte_{\text{NO}}}, &\mathbf{\Sigma}_{\Xte_{\text{NO}} \Xtr_{\text{O}}}\\
    \mathbf{\Sigma}_{\Xtr_{\text{O}} \Xte_{\text{NO}}}, & \mathbf{\Sigma}_{\Xtr_{\text{O}} \Xtr_{\text{O}}} + \frac{1}{L} \widehat{\boldsymbol{\sigma}}^2_{\mathrm{z}}
    \end{pmatrix}
    \right).
\end{equation}
Here $\restr_{\text{O}}$ is a vector of partial residuals for the current treatment tree, i.e., the difference between observed response $y$ and the predictions of all other prognostic and treatment trees. $\reste_{\text{NO}}$ is a vector of residuals we wish to predict for the non-overlap testing data. $\widehat{\boldsymbol{\sigma}}^2_{\mathrm{z}}$ is a diagonal matrix with diagonal elements equal to estimated variance $\widehat{\sigma}_0^2$ or $\widehat{\sigma}_1^2$ depending on the treatment variable of the training data and $L$ is the total number of prognostic and treatment trees.

The prediction algorithm for the treatment trees in XBCF-GP follows the same procedures as in Algorithm $\ref{alg:PFR}$ except for the difference we describe above. Specifically, we replace the out-of-range hypercube $\mathcal{B}_{lb}$ in line $4$ by the overlap area $\mathcal{B}_O$. In line $6$, the subset of training data $\Xtr_b$ is sampled from the training data $\Xtr_{\text{O}}$ from the overlap area instead of the entire training set. Then in line $7$, we choose to extrapolate the testing data $\Xte_{\text{NO}}$ in the non-overlap area for XBCF-GP. Similar to the algorithm design choices made in XBART-GP, the overlap area is defined by the $95\%$ quantiles of the treated and control data for robustness concerns.

In addition to the modification we make to the XBART-GP algorithms, we also add a tweak to the original XBCF model to ensure the extrapolation works smoothly. To ensure successful extrapolation, the GP requires a minimum amount of overlap data in the leaf nodes for providing accurate estimates. While it is uncommon for the treatment trees in XBCF to split in the non-overlap area, it could still occur in rare cases and result in suboptimal extrapolation. To address this issue, we implemented a strong prior on the treatment trees to discourage splits that do not have sufficient treatment and control data in their children nodes. The amount of overlap data required in a leaf node can be controlled through the hyperparameter {\tt Nmin}. This hyperparameter is critical as it determines the balance between the quality of the method's estimates near the overlap area boundary and its extrapolation performance in the non-overlap region. In our experiments using $500$ data points, we set {\tt Nmin} to $20$.

The XBCF-GP model combines the benefits of the Bayesian Causal Forest and Gaussian Process. XBCF excels at accurately and efficiently estimating homogeneous and heterogeneous treatment effects on large datasets with many covariates, provided the positivity assumption holds. The GP component, applied on a per-leaf node basis, enables extrapolation of treatment effects in regions where the positivity assumption is violated by leveraging the most relevant covariates in the surrounding area.

Our method provides a more effective way to assess the uncertainty of the estimated treatment effect in cases where the positivity assumption is violated. By using the local Gaussian process, our uncertainty estimation takes into account the distance between the testing data and the overlap region. This means that if the testing data is significantly distant from the overlap region, XBCF-GP will reflect this uncertainty through wider prediction intervals.

\subsection{Simulation study} \label{sec:xbcf-sim}
This section compares prediction interval coverage and length under the setting of causal inference when the positivity assumption is violated. To demonstrate the performance of XBCF-GP, we compare it with a set of alternative extrapolation approaches proposed by \cite{nethery2019estimating}. The baseline methods we use are summarized as follows: 
\begin{itemize}
    \item{\textbf{SBART}} Also known as BART-Stratified. Fit separate BART models for the treated and the control group, then estimate the treatment effect as the difference between the expected value of the treated and the control models.
    \item{\textbf{UBART}} Untrimmed BART is implemented by \cite{nethery2019estimating}, in which a single BART model is fitted with covariates, treatment variable, and estimated propensity score. The treatment effect is estimated by predicting the potential outcomes with the posterior predictive distributions. 
    \item{\textbf{BART+SPL}} \cite{nethery2019estimating} proposed this method to estimate the overlap area with BART and the non-overlap area with a spline model. The region of overlap is defined by propensity score with recommended parameters $a = 0.1$ and $b = 7$.
\end{itemize}

\begin{figure}[h!]
    \centering
    \begin{tabular}{cccc}
    \includegraphics[width=0.32\textwidth]{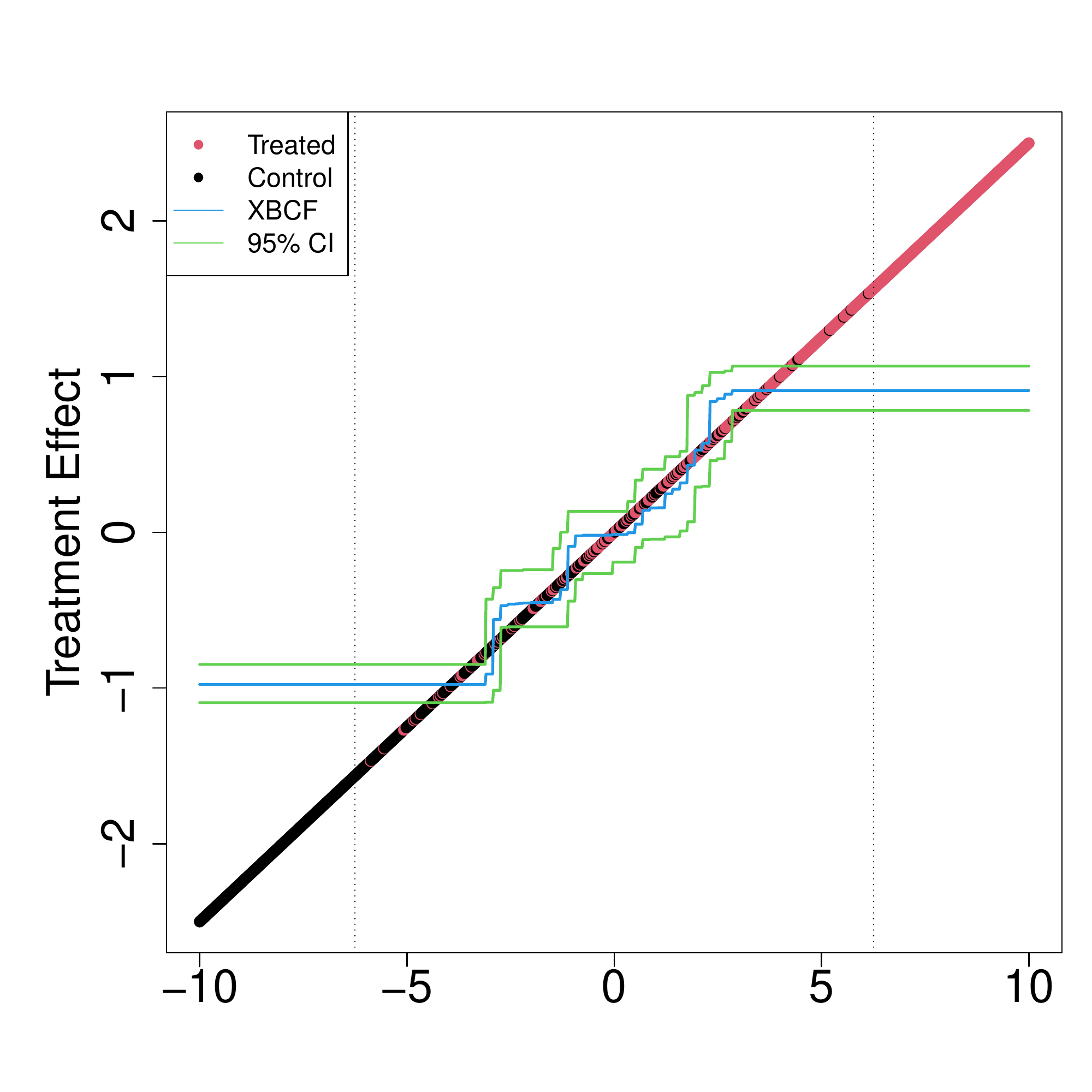}  &
    \includegraphics[width=0.32\textwidth]{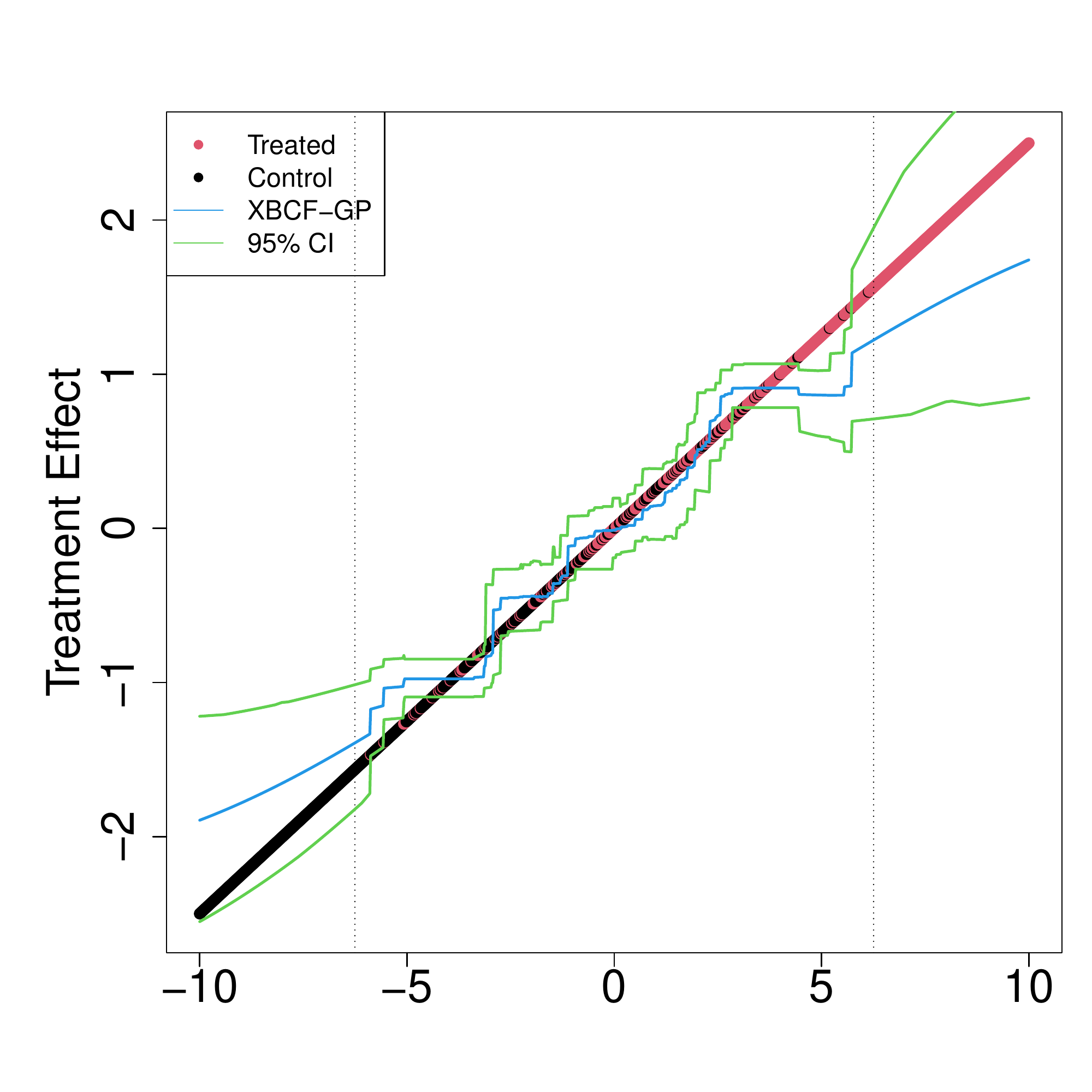}  \\
    \text{(a) XBCF}  & \text{(b) XBCF-GP}  \\[6pt]
    \end{tabular}
    \begin{tabular}{cccc}
    \includegraphics[width=0.32\textwidth]{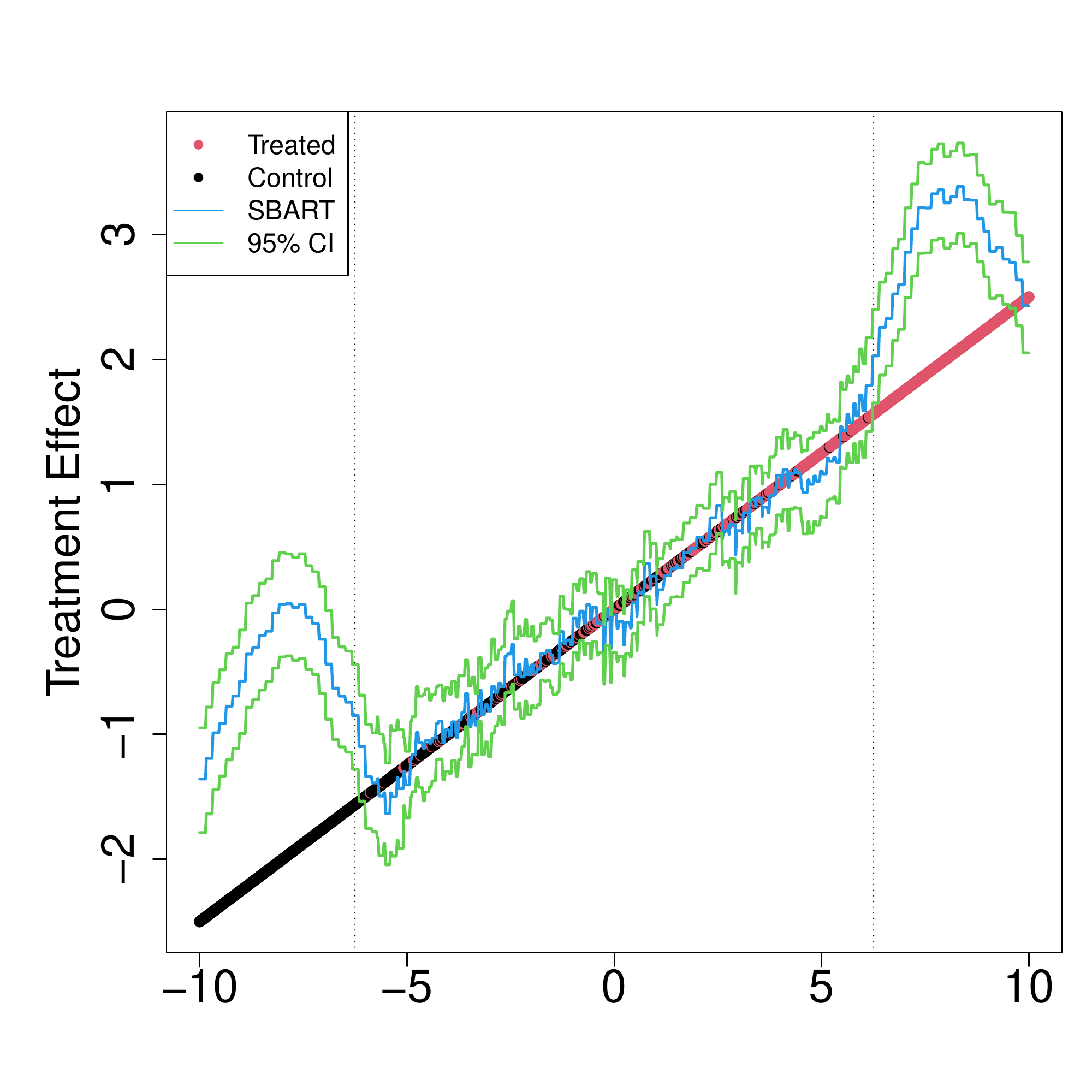} &
    \includegraphics[width=0.32\textwidth]{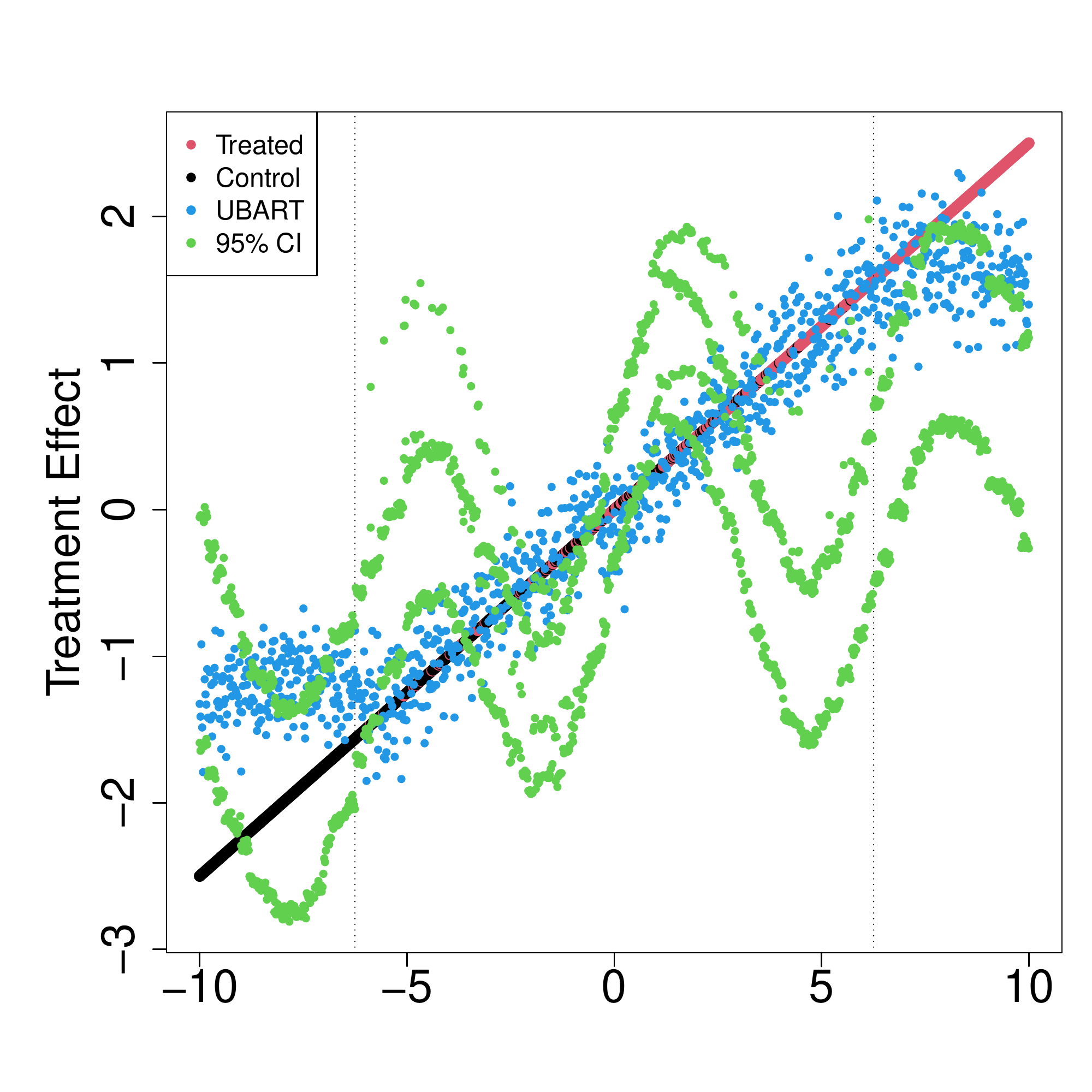} &
    \includegraphics[width=0.32\textwidth]{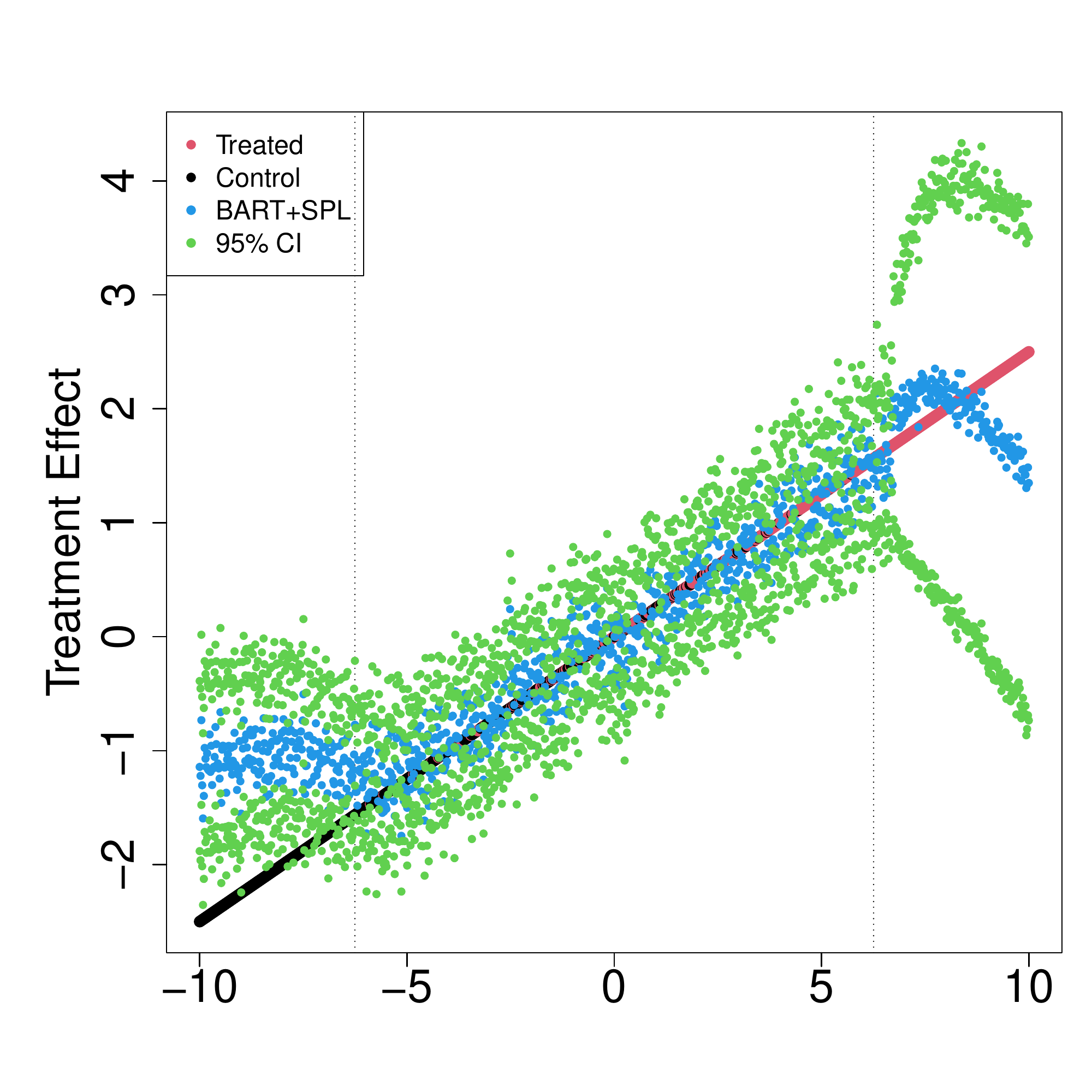} \\
    \text{(c) SBART}  & \text{(d) UBART} & \text{(e) BART+SPL}  \\[6pt]
    \end{tabular}
 
    \caption{An example of treatment effect estimation with positivity violation on all baseline methods. The two vertical lines indicate the boundaries of the overlap area, with the control group heavily distributed on the left side and the treatment group distributed on the right side. The straight line represents the true treatment effect, while the curved lines/dots in the center denote the estimated treatment effect for each method. The lower and upper curved lines/dots show their respective $95\%$ predictive intervals.}
    \label{fig:xbcf_demo_full}
\end{figure}

Figure \ref{fig:xbcf_demo_full} illustrates the performance of XBCF-GP and other baseline methods on a one-dimensional toy dataset. The independent variable $x$ is uniformly drawn from the range $[-10,10]$. The prognostic function is sine, and the treatment effect is $0.25x$. The probability for a sample being treated is $\pi(x) = \max\{0, \min\{1, 0.08x + 0.5\}\}$. Let $f(x) = \sin(x) + (0.25x)z$ denotes the true function with treatment variable $z \sim \text{Bern}(\pi(x))$. The response values $y$ are generated as $y = f(x) + \epsilon$, where $\epsilon \sim N(0, 0.2 \times\text{sd}(f))$ with the standard deviation $\text{sd}(f)$ taken over the dataset. The fitted models in the example use the same parameter setup as in the simulation study, which will be specified shortly.

The figure demonstrates the performance of different models in estimating CATE. The first three methods (XBCF, XBCF-GP, and SBART) are relatively robust to noise. However, the last two methods (UBART and BART+SPL), which directly estimate CATE using the response variable, introduce significant noise in the estimates. All three BART-based methods are susceptible to the influence of the prognostic structure, especially in the non-overlap area. XBCF dodges the bullet by setting up a separate BART forest to control for the prognostic effect, but it cannot extrapolate properly in positivity violation. XBCF-GP inherits the well-regularized structure from XBCF and extrapolates the treatment effect estimations nicely and smoothly.

For a more thorough comparison among these methods, we adopt the same data generating process used in \cite{hahn2020bayesian} and \cite{nikolay2022} but modify the propensity function to create the positivity violation scenario. The covariate vector $\x$ consists of five variables. The first three (denoted as $x_1, x_2, x_3$) are generated from the standard normal distribution. $x_4$ is a binary variable, and $x_5$ is an unordered categorical variable with levels denoted as $1, 2, 3$. The prognostic function can be either linear or nonlinear:
\[
    \mu(\x) = \begin{cases} 
      1 + g(x_4) + x_1 x_3 \quad & \text{linear} \\
      -6 + g(x_4) +6 |x_3 - 1 | \quad & \text{nonlinear,}
   \end{cases}
\]
where $g(1) = 2$, $g(2) = -1$ and $g(3) = -4$. The treatment effect can be either homogeneous or heterogeneous:
\[
\tau(\x) = \begin{cases}
    3 \quad & \text{homogeneous} \\
    1 + 2x_2 x_5 \quad & \text{heterogeneous.} 
\end{cases}
\]

In the simulation studies, we adjust the propensity function such that roughly $20\%$ of the data has a propensity score of $1$, $20\%$ of the data has a propensity score of $0$, and the rest of the data has a propensity score that is between $0$ and $1$. The propensity function $\pi(\x)$ is given by
\begin{equation}
   \pi(\x) = \max \{0, \min \{1, h(\x)\} \}, \quad h(\x) = 1.1 \Phi(3\mu(\x)/s - 0.5x_1 - c) - 0.15 + u_i/10,
\end{equation}
where $s$ is the standard deviation of $\mu(\x)$ taken over the observed samples and $u_i \sim \text{Uniform}(0,1)$. We pick the value $c = 0$ for the linear prognostic function and $c = 3$ for the non-linear one. For any propensity score greater than $1$ or less than $0$, we set it to be $1$ or $0$, respectively. In addition, we add a Gaussian noise $\epsilon \sim N(0, 0.5 \times \text{sd}(\mu(\x) + \tau(\x)\mathrm{z}))$ with the standard deviation taken over the samples to generate the response values. 

\begin{table}[h! ]
	\small
	\def\arraystretch{1}
	\centering 
	\begin{tabular}{c|ccc|ccc|c}
		\toprule
		\multirow{2}{*}{Method} & \multicolumn{3}{c|}{ATE}  & \multicolumn{3}{c|}{CATE} & \multirow{2}{*}{Time} \\
		         & RMSE  & Coverage & I.L.  & RMSE  & Coverage & I.L.   &       \\
		\toprule
		\multicolumn{8}{c}{Linear Homogeneous}                                                                \\
		\hline
        XBCF     & 0.369    & 0.600          & 0.927   & 0.474     & 0.944         & 1.670     & 0.998   \\
        XBCF-GP & 0.347    & 0.600          & 0.903   & 0.467     & 0.953         & 1.736    & 2.026   \\
        SBART    & 1.148    & 0.000            & 0.945   & 1.511     & 0.726         & 3.216    & 7.155   \\
        UBART    & 0.290     & 1.000            & 1.180    & 1.299     & 0.995         & 5.684    & 6.338   \\
        BART+SPL  & 0.987    & 1.000            & 3.713   & 1.725     & 0.987         & 13.806   & 145.865 \\
		\toprule
		\multicolumn{8}{c}{Linear Heterogeneous}                                                              \\
		\hline
		XBCF     & 0.630     & 0.800          & 1.579   & 1.773     & 0.755         & 3.566    & 0.978   \\
        XBCF-GP & 0.601    & 0.800          & 1.671   & 1.671     & 0.774         & 3.720     & 2.707   \\
        SBART    & 1.241    & 0.000         & 1.523   & 1.979     & 0.858         & 5.589    & 7.088   \\
        UBART   & 0.393    & 1.000       & 2.071   & 2.213     & 0.851         & 9.032    & 6.344   \\
        BART+SPL  & 0.958    & 1.000          & 6.485   & 2.748     & 0.96          & 24.523   & 158.565 \\
		\toprule
		\multicolumn{8}{c}{Non-linear Homogeneous}                                                            \\
		\hline
		XBCF     & 0.786    & 0.800       & 2.254   & 0.627     & 0.954         & 2.709    & 0.910    \\
        XBCF-GP & 0.687    & 0.800       & 2.352   & 0.575     & 0.960         & 2.924    & 1.827   \\
        SBART    & 2.370     & 0.000         & 2.586   & 3.288     & 0.802         & 7.821    & 7.095   \\
        UBART    & 0.556    & 1.000        & 2.704   & 2.819     & 0.929         & 12.867   & 6.306   \\
        BART+SPL  & 1.439    & 1.000         & 8.342   & 3.476     & 0.989         & 27.899   & 159.738 \\
		\toprule
		\multicolumn{8}{c}{Non-linear Heterogeneous}                                                          \\
		\hline
        XBCF     & 1.219    & 0.600       & 2.456   & 2.424     & 0.759         & 5.147    & 0.940    \\
        XBCF-GP & 1.20      & 0.600       & 2.628   & 2.354     & 0.792         & 5.576    & 2.270   \\
        SBART    & 2.931    & 0.000         & 2.708   & 4.051     & 0.780         & 8.783    & 7.134   \\
        UBART    & 0.801    & 0.800        & 2.718   & 3.297     & 0.840        & 14.116   & 6.326   \\
        BART+SPL  & 2.166    & 0.800         & 5.688   & 4.458     & 0.970         & 24.641   & 161.630  \\
		\bottomrule
	\end{tabular}
	\caption{\label{tab:cate} Results of root mean square error (RMSE), interval coverage (Coverage), interval length (I.L.), and running time (in seconds) for ATE and CATE estimators with different combinations of treatment term and prognostic term types. The sample size is 500. } 
\end{table}

To gauge the performance of XBCF-GP across overlap and non-overlap regions, we evaluate the average treatment effect (ATE) and conditional average treatment effect (CATE) on three basic metrics: average root mean square error, coverage, and average interval length. In addition, we also compare the time efficiency of the proposed method to baseline methods. We average the results from $20$ independent replications for each scenario with $n = 500$ samples.

For methods that rely on the propensity score, we use the propensity score estimated by a XBART Multinomial model with $2$ classes and $100$ iterations instead of the true propensity scores. For XBCF-GP, we set hyper-parameter $\theta = 0.1$ and $\tau = \widehat{\sigma}^2 / L_{\tau}$, where $\widehat{\sigma}^2$ is the estimated variance and $L_{\tau} = 20$ is the default number of treatment trees in the XBCF model. We choose a different scale parameter $\tau_{gp}$ from XBART-GP as the variance of the treatment effect tends to be smaller than the total variance of the observed data in most cases. As for the other baseline methods, including XBCF, we use their default parameters for all scenarios. 

The experiments were conducted in \texttt{R} 4.1.3 on the same machine described in the previous sections. 

Table \ref{tab:cate} summarizes the simulation results of various methods. We focus our discussion here on CATE estimation. Both XBCF methods provide the best CATE estimation in all cases and deliver competitive CATE coverage with tighter intervals than other methods. XBCF-GP can improve both the accuracy and coverage of ATE and CATE estimation based on its base model. Furthermore, we notice that the efficiency of XBCF-GP, inherited from XBART, is substantially faster than other BART-based methods. 

\section{Discussion}
The Gaussian process extrapolation technique developed here has several advantages over standard BART extrapolation, as implemented in the most widely available software. First, whereas BART extrapolates a constant function beyond the range of the training data by fusing a Gaussian process at each leaf, ``local'' Gaussian process extrapolation is possible. This local GP approach automatically incorporates variable relevance, using only variables that are used as splitting rules in a given tree. Because extrapolation occurs at each tree, our procedure also generally improves prediction: GP prediction at {\em local} exterior points imposes additional smoothness even at points within the training data's convex hull.

Second,  BART's prediction intervals are dominated by the residual variance parameter value, which implies that the interval width is nearly constant at any exterior point. By contrast, the local GP extrapolation method implies interval widths that expand quite rapidly the further away a point is from the training data, a desirable property that is directly inherited from the Gaussian process.

Third, the local GP approach requires no additional training time compared to a regular BART model fit. Instead, the extrapolation component is a direct modification of the posterior predictive distribution defined in terms of BART posterior samples and a specified covariance function. Broadly, our approach grew out of attempts to answer the question: how would one want to use a known BART model that fits well at interpolation points to define an extrapolation model? Our proposed solution is to graft a Gaussian process to each leaf node for the purpose of extrapolation only; posterior uncertainty in the BART model parameters is then propagated by averaging over posterior samples. The general philosophy behind this approach is that there is no need to change the likelihood model if BART adequately fits the training data, but conversely, there is no need to stick to the pure BART model for points far from the training data. Extrapolation always requires a leap of faith, and there is no guarantee that any extrapolation method will succeed, but we argue that there are conceptual and computational advantages to disentangling the extrapolation model from the interpolation model.

Finally, we demonstrate how to apply local GP extrapolation of BART models to the problem of imperfect overlap in treatment effect estimation, a common practical challenge in applied causal inference. By applying the local GP extrapolation technique to the Bayesian causal forest model, violations of the positivity assumption in practice can be handled organically. Specifically, we are able to extrapolate the treatment effect surface directly rather than composing it as the difference between two separate extrapolations (one for each treatment arm). Our simulation studies demonstrate that XBCF-GP produces treatment effect estimates with good coverage on a variety of data-generating processes. These tools will help other researchers to draw more reliable inferences from applied data which frequently exhibit imperfect overlap.

\paragraph{Acknowledgement} Jingyu He gratefully acknowledges funding for this project fully supported by the Research Grants Council of the Hong Kong Special Administrative Region, China (Project No. CityU 21504921). 


\section{Supplementary Materials}
\begin{description}
\item[R-package for XBART-GP:] R-package "XBART" is adapted from the original XBART model (\url{https://github.com/JingyuHe/XBART}) \citep{he2019xbart, he2021stochastic} and contains code to perform local Gaussian process extrapolation on trained XBART model described in this artical. Please read file README contained in the zip file for installation instruction. (XBART.zip, zip archive)
\item[Python-package for XBART-GP:] Python-package "XBART"is adapted from the original XBART model (\url{https://github.com/JingyuHe/XBART}) \citep{he2019xbart, he2021stochastic} and contains code to perform local Gaussian process extrapolation on trained XBART model described in Python. Please read file README contained in the zip file for installation instruction. (XBART-python.zip, zip archive)
\item[R-package for XBCF-GP:] R-package "XBCF" is adapted from the original XBCF model (\url{https://github.com/socket778/XBCF}) \citep{nikolay2022} and contains code to perform local Gaussian process extrapolation on trained XBCF model described in the artical. Please read file README contained in the zip file for installation instruction. (XBCF.zip, zip archive)
\item[Python code for XBART-GP simulations] The supplemental files contain code to perform simulation experiments described in Section \ref{sec:sim}. (XBART-GP simulation.zip, zip archive)
\item[R code for XBCF-GP simulations] The supplemental files contain code to perform simulation experiments described in Section \ref{sec:xbcf-sim}. (XBCF-GP simulation.zip, zip archive)

\end{description}

\bibliographystyle{Chicago}
\bibliography{BART}

\newpage

\appendix
\section{Simulation results for XBART-GP and its baselines}
\subsection{Simulation results on different functions}

\begin{figure}[!ht]

\begin{subfigure}{379pt}
\centering 
\subcaption[c]{Trig + poly}
 \includegraphics[width=379pt]{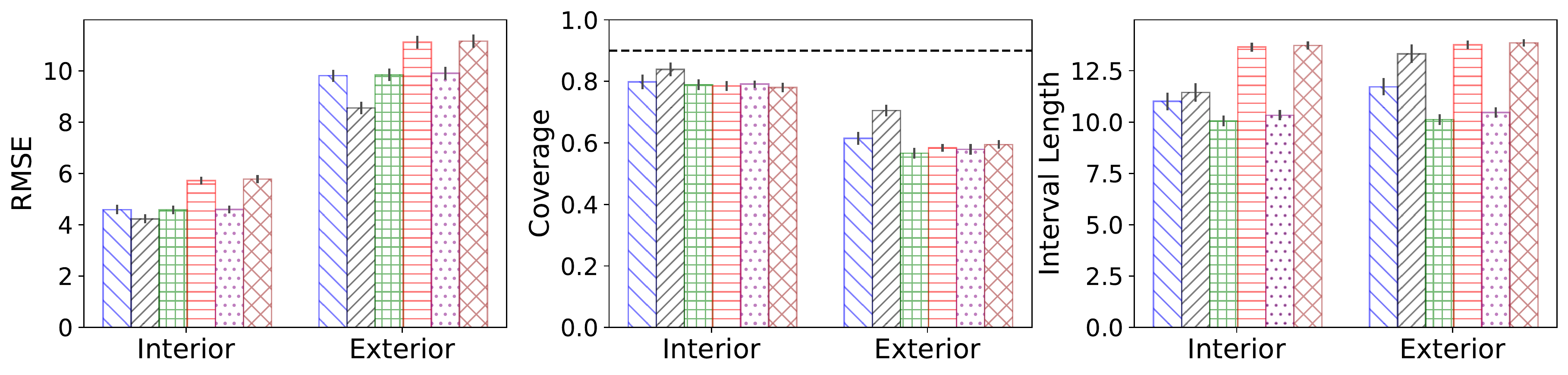}
\end{subfigure}
\centering
\begin{subfigure}{379pt}
 \subcaption[c]{Max}
 \includegraphics[width=379pt]{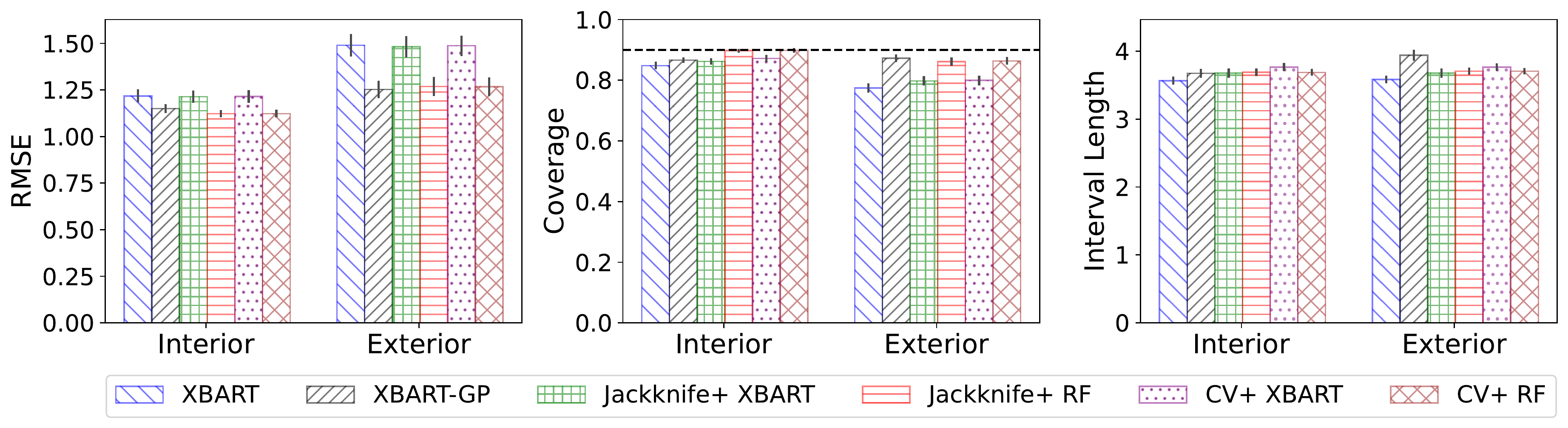}
\end{subfigure}
\caption{Visualization of simulation results in root mean square error (RMSE), interval coverage (Coverage), and interval length for interior and exterior points for tirg + poly and max functions. The horizontal dash line indicates the $90\%$ target coverage level.}
\label{fig:simulation2}
\end{figure}

\begin{table}[h!]
\centering
\small
  \begin{tabular}{c|cc|cc|cc}
    \toprule
    \multirow{2}{*}{Method} & \multicolumn{2}{c|}{RMSE} & \multicolumn{2}{c|}{Coverage} & \multicolumn{2}{c}{Interval length} \\
                     & Interior & Exterior & Interior & Exterior & Interior & Exterior \\
    \bottomrule
    \multicolumn{7}{c}{Linear}                                                                                              \\
    \hline
    XBART-GP         & 1.756    & 2.506    & 0.881    & 0.816    & 5.709    & 6.717    \\
    XBART            & 1.986    & 3.532    & 0.843    & 0.584    & 5.568    & 5.678    \\
    Jackknife+ XBART & 1.894    & 3.510    & 0.897    & 0.626    & 6.120    & 6.133    \\
    Jackknife+ RF    & 2.975    & 4.598    & 0.834    & 0.656    & 8.499    & 8.550    \\
    CV+ XBART        & 1.883    & 3.466    & 0.900    & 0.651    & 6.366    & 6.384    \\
    CV+ RF           & 3.006    & 4.635    & 0.851    & 0.652    & 8.642    & 8.702    \\
    \bottomrule
    \multicolumn{7}{c}{Single index}                                                                                        \\
    \hline
    XBART-GP         & 4.582    & 10.631   & 0.871    & 0.474    & 13.938   & 15.854   \\
    XBART            & 5.004    & 13.055   & 0.830    & 0.297    & 13.625   & 13.965   \\
    Jackknife+ XBART & 4.802    & 13.027   & 0.873    & 0.305    & 14.715   & 14.778   \\
    Jackknife+ RF    & 8.077    & 17.068   & 0.765    & 0.263    & 19.592   & 19.689   \\
    CV+ XBART        & 4.754    & 13.097   & 0.897    & 0.327    & 15.725   & 15.780   \\
    CV+ RF           & 8.165    & 17.162   & 0.764    & 0.265    & 19.880   & 19.940   \\
    \bottomrule
    \multicolumn{7}{c}{Trig + poly}                                                                                         \\
    \hline
    XBART-GP         & 4.229    & 8.549    & 0.839    & 0.705    & 11.441   & 13.322   \\
    XBART            & 4.591    & 9.805    & 0.798    & 0.615    & 11.007   & 11.718   \\
    Jackknife+ XBART & 4.577    & 9.845    & 0.789    & 0.567    & 10.054   & 10.115   \\
    Jackknife+ RF    & 5.719    & 11.110   & 0.785    & 0.584    & 13.650   & 13.764   \\
    CV+ XBART        & 4.595    & 9.904    & 0.791    & 0.579    & 10.334   & 10.470   \\
    CV+ RF           & 5.778    & 11.156   & 0.780    & 0.595    & 13.726   & 13.857   \\
    \bottomrule
    \multicolumn{7}{c}{Max}                                                                                                 \\
    \hline
    XBART-GP         & 1.150    & 1.253    & 0.866    & 0.873    & 3.672    & 3.940    \\
    XBART            & 1.218    & 1.489    & 0.848    & 0.774    & 3.564    & 3.582    \\
    Jackknife+ XBART & 1.213    & 1.482    & 0.862    & 0.798    & 3.677    & 3.678    \\
    Jackknife+ RF    & 1.123    & 1.268    & 0.898    & 0.861    & 3.689    & 3.704    \\
    CV+ XBART        & 1.215    & 1.487    & 0.872    & 0.800    & 3.765    & 3.764    \\
    CV+ RF           & 1.123    & 1.267    & 0.899    & 0.864    & 3.687    & 3.705    \\
    \bottomrule
  \end{tabular}
  \caption{Results of simulation comparing various approach on prediction interval. Columns are mean square error (RMSE), interval coverage (Coverage), and interval length (I.L.) on interior and exterior points for $4$ different data generating processes.}
  \label{tab:simulation}
\end{table}

Figure \ref{fig:simulation2} and Table \ref{tab:simulation} demonstrate the performance of XBART-GP versus other baseline methods. The results on the two functions, trig + poly and max, are consistent with the other two presented in the paper. It shows XBART-GP's ability to provide the most accurate prediction on both in-range and out-of-the-range data with almost desired coverage and slightly wider prediction intervals on various data generating processes.

\subsection{Simulation results on increasing sample size}
\begin{figure}[!ht]
 \centering
 \includegraphics[width=379pt]{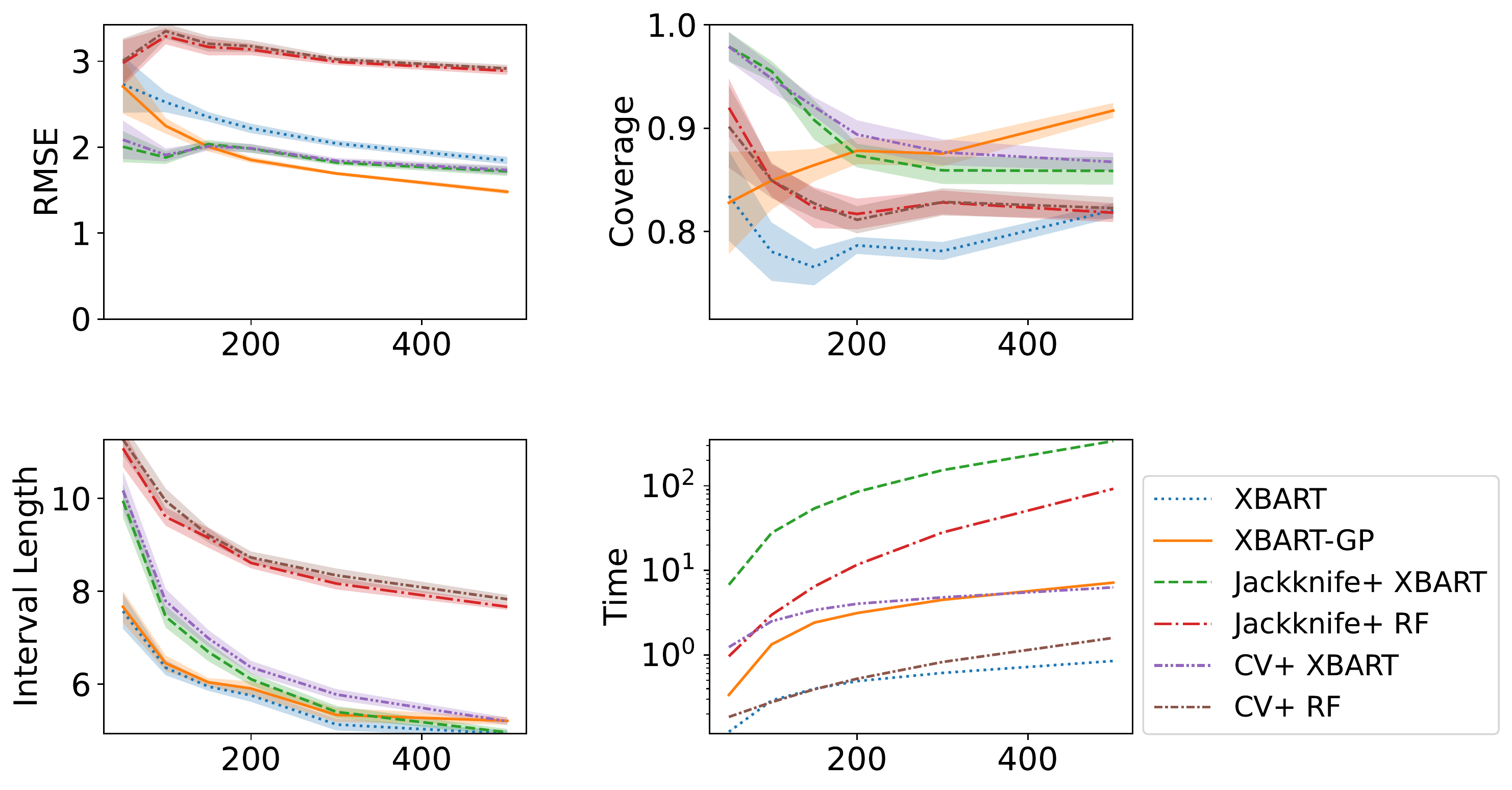}
\caption{Results of simulation comparing various approach on prediction interval. Average RMSE, coverage, interval length on interior points, and running time at $90\%$ level for all methods over $10$ independent trials with increasing sample size from $50$ to $500$.  }
\label{fig:time_interior}
\end{figure}

For the interior points, XBART-based methods produce smaller RMSE than random forest methods. XBART-GP has the smallest RMSE overall. XBART and its Gaussian process extrapolation alternative have lower coverage than Jackknife methods when the sample size is minimal. However, XBART-GP outperforms other methods on interior points as the sample size increases. XBART and XBART-GP produce the shortest intervals. The intervals of Jackknife+ XBART and CV+ XBART converge to XBART as the sample size increases.

\end{document}